\documentclass[journal]{IEEEtran}
\usepackage{amsfonts}
\usepackage{graphicx}
\usepackage{epstopdf}
\usepackage{amsmath}
\usepackage{amssymb}
\usepackage{algorithmic}
\usepackage{array}
\usepackage{url}
\usepackage{cite}
\usepackage{subfigure}
\usepackage{graphicx}
\usepackage{booktabs}
\usepackage{mathtools}
\usepackage{algorithm}
\usepackage{color}



\begin{document}

\title{Designing FDA Radars Robust to Contaminated Shared Spectra}
\author{ 
Wenkai Jia, Andreas Jakobsson, {\em Senior Member, IEEE}, Wen-Qin Wang, {\em Senior Member, IEEE}
	
\thanks{This work was supported in part by the National Natural Science Foundation of China under grant no. 62171092, and in part by the Swedish SRA ESSENCE (grant no. 2020 6:2).
	
	Wenkai Jia and Wen-Qin Wang are School of Information and Communication Engineering, University of Electronic Science and Technology of China, Chengdu, 611731, P. R. China. E-mail: wenkai.jia@matstat.lu.se.
	
	Andreas Jakobsson is with the Division of Mathematical Statistics, Center for Mathematical Sciences, Lund University, SE-22100 Lund, Sweden. E-mail: andreas.jakobsson@matstat.lu.se.
}
}



\maketitle

\begin{abstract}
This paper considers the problem of jointly designing the transmit waveforms and weights for a frequency diverse array (FDA) in a spectrally congested environment in which unintentional spectral interferences exist.
Exploiting the properties of the interference signal induced by the processing of the multi-channel mixing and low-pass filtering FDA receiver, the interference covariance matrix structure is derived.
With this, the receive weights are formed using the minimum variance distortionless response (MVDR) method for interference cancellation.
Owing to the fact that the resulting output signal-to-interference-plus-noise ratio (SINR) is a function of the transmit waveforms and weights, as well as due to the ever-greater competition for the finite available spectrum, a joint design scheme for the FDA transmit weights and the spectrally compatible waveforms is proposed to efficiently use the available spectrum while maintaining a sufficient receive SINR.
The performance of the proposed technique is verified using numerical simulations in terms of the achievable SINR, spectral compatibility, as well as several aspects of the synthesized waveforms.

\end{abstract}


\section{Introduction}
\IEEEPARstart{F}{requency} diverse array (FDA) systems make use of an array configuration scheme \cite{1631800}, which employs a small frequency increment across the adjacent array elements \cite{7740083}.
Such systems are capable of producing a range-angle-time dependent transmit beampattern \cite{8321488,6376087,8550659,6509929} that allows for range-ambiguous clutter suppression \cite{7181636}, deceptive mainlobe interference suppression \cite{9161264}, low probability of identification (LPI) \cite{9440812}, joint range and angle estimation \cite{6630081,6737322}, radio frequency (RF) stealth \cite{7362565,7422108}, and high-resolution synthetic aperture radar (SAR) imaging \cite{7559704}, and have for these reasons attracted notable attention over the past decade.

As an example of the potential of using an FDA, a study of the Doppler-spreading (DS) effect in the Doppler domain of the FDA echo signal reflected by a moving target was reported in \cite{gui2021fda}, showing that this DS effect may allow for resolving Doppler ambiguity and can therefore be used for target detection in mainlobe clutter.
Moreover, the cognitive FDA for target tracking was investigated in \cite{gui2021cognitive} by adaptively designing the transmit weight matrix according to the available prior knowledge at each transmission.
Additionally, considering the time-variance property of an FDA transmission, as discussed in \cite{4250348,6786333}, Gui \emph{et al.} proposed a multi-channel matched filtering receiver structure to process pulsed-FDA signals \cite{8074796}.
Although computationally efficient, the resulting receiver would suffer serious performance loss in the presence of signal-dependent interference, as it only maximizes the signal-to-noise ratio (SNR), but not the signal-to-clutter ratio (SCR).
To alleviate this drawback, we recently presented a multi-channel mixing and low-pass filtering receiver maximizing output signal-to-interference-plus-noise ratio (SINR) by designing optimal transmit waveforms to this effect in \cite{1111111}.

However, the aforementioned work are established on the assumption that the radar is the exclusive user of its allocated frequency spectrum and that the statistical characteristics of the FDA receiver output interference signals are similar to those of a conventional multiple-input multiple-output (MIMO) radar.
In reality, the demand for wireless access by, e.g., mobile communication and navigation systems, is steadily increasing, thereby eroding the spectrum allocation assigned to the radar community \cite{6967722}.
Furthermore, most such systems have a need for ever greater bandwidth and with the disappearance of guard bands, this has become a severe and growing problem \cite{6170864}.
For instance, the Global Positioning System (GPS) and Bluetooth operate in the radar $L$- and $S$-bands. At higher frequencies, the services will also include microwave and satellite communications links.
Since the bandwidth of a radar pulse determines its sensing capability, the narrower frequency bands available might result in false detections \cite{richards2014fundamentals}.
For FDA, a greater bandwidth has to be satisfied for the reason that the used frequency increment widens the used spectrum, especially for FDA-MIMO, where the frequency increment is larger than the bandwidth of the transmitted baseband signal \cite{7084678,6404099,8049352}.
At the same time, in the presence of radar interference, the ability of a communication receiver to recover the transmitted symbol will be severely degraded \cite{8828016}.
Therefore, a better solution is to construct the systems to allow different services to operate within the same frequency bands with a tolerable level of distortion.

In addition to the above discussion, the statistical characteristics of interferences must be considered when analyzing and processing FDA output signals. 
In \cite{8074796}, a noise model is derived to allow for the fact that the FDA output noise does not obey the independence assumption after passing through the multi-channel receiver, in contrast to that of a MIMO or a phased-array (PA).
Likewise, the interference model needs to be modified to ensure that the appropriate statistical characteristics are provided to obtain a more accurate estimate of the interference-plus-noise covariance matrix.
Meanwhile, modern digital technology enables the creation of precise radar waveforms with spectral nulls at particular frequency bands.
This also allows for an optimal solution wherein the FDA waveforms may be designed specifically for the spectrally congested environment in which they will operate, taking into account the structure of the present interference signals.

There are many excellent reviews in the literature dealing with the design of PA or MIMO radar in a spectral congestion scenario (see, e.g., \cite{8579200,6850145,9052442,8356676}). 
However, unlike PA, which transmits a single waveform \cite{7838312}, and MIMO, where the multiple waveforms will cause a dependency of the waveform spectral distribution on the spatial direction \cite{8356676}, FDA contains multiple waveforms with large frequency increments, resulting in an omnidirectional distribution of electromagnetic energy.
To the best of our knowledge, no well working solution to this problem has been proposed and how one should design spectrally compatible waveforms for FDA is still a problem that merits further attention.
Not limited to waveform design, in recent years, several approaches have been proposed to deal with the problem of spectrum congestion and to allow for a more efficient spectrum usage \cite{cardinali2007multipath,6170864,6967722}.
The advantage of the waveform design approach is that, relying on real-time spectrum occupancy sensing, the waveform can be dynamically selected based on changing conditions, thereby controlling its impact on other compatible systems.
In addition, the underlying optimization process can also benefit from multiple design degrees of freedom to further enhance the radar performance.

Rather than assuming that no other radiators will occupy the same part of the spectrum as the current radar application, we will in  this work take into account for the increasingly crowded spectrum, proposing an FDA system that allows for the presence of other applications in the same spectral band that is used by the radar.
The main contributions and novelties of this work are summarized as follows.
\begin{itemize}
	\item[i)]  Employing the multi-channel mixing and low-pass filtering-based FDA receiver presented in \cite{1111111}, the transmit waveforms can be extracted from the signals at the receiver output end. This allows the FDA spectrum compatibility constraint and the interference-plus-noise covariance matrix to be derived. Since the requirements for FDA waveform separation in the receiver requires that a constant energy for each waveform must be guaranteed \cite{1111111}, the transmit weights are designed to achieve less mutual interference than fixed weights.
	\item[ii)] Aiming at minimizing the mutual interference induced by frequency overlaid systems, the joint design of the FDA transmit waveforms and weights is formulated as an optimization problem with multiple non-convex constraints, using the output SINR as the objective function to measure the system performance.
	In order to solve the resulting non-convex problem, an iterative algorithm based on the semidefinite relaxation (SDR) technique presented in \cite{5447068} is introduced. 
\end{itemize}
Various numerical simulations are carried out to demonstrate the performance of the proposed algorithm from the point of achievable SINR and spectral compatibility as well as with respect to several aspects of the radar environment.
It is found that the design scheme based on waveform optimization not only maximizes the output SINR, but also enables finer control over the operating spectrum of the system.

The remainder of the paper is organized as follows.
In Section \ref{sec2}, the assumed signal model and the statistical analysis of the potential interferences are presented. Next, we formulate the optimization problem maximizing the output SINR subject to the derived FDA spectrum compatibility constraint in Section \ref{sec3}. In Section \ref{sec4}, an iterative method based on an SDR technique is developed to solve the resulting optimization problem. Simulation results are given in Section \ref{sec5} and, finally, our conclusions are drawn in Section \ref{sec6}.

\section{Signal Model}
\label{sec2}
\subsection{Receive Signal Model}
\begin{figure}[htb]
	\centering
	\includegraphics[width=0.45\textwidth]{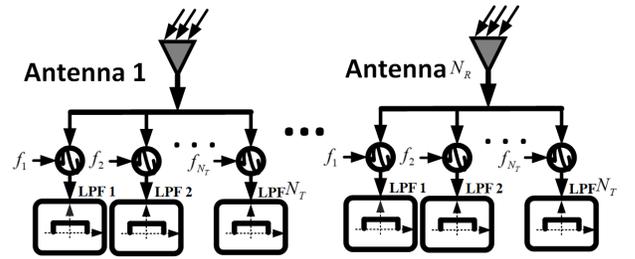}
	\caption{The multi-channel mixing and low-pass filtering-based FDA receiver.}
	\label{Fi1}	
\end{figure}
Consider a uniform linear FDA, with $N_T$ transmit antennas, each of which radiates a different waveform ${s_m}\left( t \right),m = 1,2,...,{N_T}$. 
The transmitted radio frequency FDA signal may be expressed as
\begin{equation}
x\left( t \right) = \sum\limits_{m = 1}^{{N_T}} {{w_m}{s_m}\left( t \right){e^{j2\pi {f_m}t}}}
\end{equation}
where ${f_m} = {f_c} + \left( {m - 1} \right)\Delta f$, with $\Delta f$ being the frequency increment, and $w_m$ denoting the carrier frequency and the transmit weight of the $m$-th transmit antenna, respectively.
Thus, the synthesized signal from a given far-field point target location $\left( {{\theta _t},{r_t}} \right)$, where $r_t$ and $\theta_t$ represent the slant range and azimuth angle, respectively, can be expressed as \cite{8074796}
\begin{align}
\kern 2pt &y\left( t \right) \nonumber \\
& =\sum\limits_{m=1}^{{{N}_{T}}}{{{w}_{m}}{{s}_{m}}\left( t-\frac{{{r}_{t}}-m{{d}_{t}}\sin {{\theta }_{t}}}{c} \right){{e}^{j2\pi {{f}_{m}}\left( t-\frac{{{r}_{t}}-m{{d}_{t}}\sin {{\theta }_{t}}}{c} \right)}}}  \nonumber \\
& \approx \sum\limits_{m=1}^{{{N}_{T}}}{{{w}_{m}}{{s}_{m}}\left( t-\frac{{{r}_{t}}}{c} \right){{e}^{j2\pi {{f}_{m}}\left( t-\frac{{{r}_{t}}}{c} \right)}}{{e}^{j2\pi \frac{m{{d}_{t}}\sin {{\theta }_{t}}}{\lambda }}}} \nonumber \\
& ={{\left[ \mathbf{w}\odot {{\mathbf{a}}_{T}}\left( {{\theta }_{t}} \right) \right]}^{T}}\mathbf{\tilde{s}}\left( t-\frac{{{r}_{t}}}{c} \right)
\end{align}
where ${\bf{w}} = {[{w_1},{w_2},...,{w_{{N_T}}}]^T}$ denotes the transmit weight vector, with $^T$ being the
transpose operator, ${\bf{\tilde s}}\left( t \right) = {\bf{s}}\left( {t - \frac{r}{c}} \right) \odot {\bf{e}}\left( t \right)$, with 
\begin{equation}
{\bf{s}}\left( t \right) = {\left[ {{s_1}\left( t \right),{s_2}\left( t \right),...,{s_{{N_T}}}\left( t \right)} \right]^T}
\end{equation}
and 
\begin{equation}
{\bf{e}}\left( t \right) = {\left[ {{e^{j2\pi {f_c}t}},{e^{j2\pi \left( {{f_c} + \Delta f} \right)t}},...,{e^{j2\pi \left( {{f_c} + \left( {{N_T} - 1} \right)\Delta f} \right)t}}} \right]^T},
\end{equation}
being the transmit waveform vector and the carrier vector, respectively, $\lambda  = \frac{c}{{{f_c}}}$ the reference wavelength, with $c$ denoting the speed of light,
\begin{equation}
{{\bf{a}}_{\mathop{T}\nolimits} }\left( {{\theta _t}} \right) = {\left[ {1,{e^{j2\pi {d_t}\frac{{\sin {\theta _t}}}{\lambda }}},...,{e^{j2\pi \left( {{N_T} - 1} \right){d_t}\frac{{\sin {\theta _t}}}{\lambda }}}} \right]^T},
\end{equation}
with ${{d}_{t}}=\frac{\lambda }{2}$ being the inter-element spacing of the transmit array, and $ \odot $ is the Hadamard product operator.
It is worth noting that the transmit power allocation is enabled by coupling of the transmit weight vector ${\bf{w}}$ with the transmit waveform vector ${\bf{s}}\left( t \right)$.
Here, we assume that the $N_R$ receive antennas are co-located with the transmit antennas, at half-wavelength intervals, ${{d}_{r}}=\frac{\lambda }{2}$. Using the multi-channel mixing and low-pass filtering-based FDA receiver \cite{1111111}, as shown in Fig. \ref{Fi1}, where a reflected signal is first mixed by the multi-channel mixers with local carrier frequencies $\left\{ {{f_m}} \right\}_{m = 1}^{{N_T}}$ and subsequently processed by the lowpass filter, the filtered output matrix may be obtained as 
\begin{align}
&{\mathbf{\bar{R}}}\left( t \right) \nonumber \\
& = \left[ {{{\bf{r}}_1}\left( t \right),{{\bf{r}}_2}\left( t \right),...,{{\bf{r}}_{{N_R}}}\left( t \right)} \right]  \nonumber \\
& = \varsigma \left( {{r_t},{\theta _t}} \right){\mathop{\rm diag}\nolimits} \left\{ {{\bf{w}} \odot {{\bf{a}}_{\mathop{T}\nolimits} }\left( {{r_t},{\theta _t}} \right)} \right\}{\bf{s}}\left( {t - \frac{{2{r_t}}}{c}} \right){\bf{b}}_R^T\left( {{\theta _t}} \right) \nonumber \\
& = \varsigma \left( {{r_t},{\theta _t}} \right){\mathop{\rm diag}\nolimits} \left\{ {{\bf{s}}\left( {t - \frac{{2{r_t}}}{c}} \right) \odot {{\bf{a}}_{\mathop{T}\nolimits} }\left( {{r_t},{\theta _t}} \right)} \right\}{\bf{wb}}_R^T\left( {{\theta _t}} \right)
\end{align}
where $\varsigma \left( {{r_t},{\theta _t}} \right)$ is the target complex reflection coefficient, $\text{diag}\left\{ \mathbf{w}\odot {{\mathbf{a}}_{T}}\left( {{r}_{t}},{{\theta }_{t}} \right) \right\}$ denotes the diagonal
matrix with entries formed by $\mathbf{w}\odot {{\mathbf{a}}_{T}}\left( {{r}_{t}},{{\theta }_{t}} \right)$,
\begin{equation}
{{\bf{b}}_R}\left( {{\theta _t}} \right) = {\left[ {1,{e^{j2\pi {d_r}\frac{{\sin {\theta _t}}}{\lambda }}},...,{e^{j2\pi \left( {{N_R} - 1} \right){d_r}\frac{{\sin {\theta _t}}}{\lambda }}}} \right]^T}
\end{equation}
denotes the FDA angle-dependent receive steering vector, and ${{\bf{r}}_n}\left( t \right)$ the filtered output of the $n$-th receive antenna, given by
\begin{equation}
\begin{array}{l}
{{\bf{r}}_n}\left( t \right) = \varsigma \left( {{r_t},{\theta _t}} \right){e^{j2\pi \frac{{\left( {n - 1} \right){d_r}\sin {\theta _t}}}{\lambda }}}{\mathop{\rm diag}\nolimits} \left\{ {{\bf{w}} \odot {{\bf{a}}_{\mathop{T}\nolimits} }\left( {{r_t},{\theta _t}} \right)} \right\}\\
\kern 37pt \cdot {\bf{s}}\left( {t - \frac{{2{r_t}}}{c}} \right)
\end{array}
\end{equation}
with 
\begin{equation}
\begin{aligned}
& {{\mathbf{a}}_{T}}\left( {{r}_{t}},{{\theta }_{t}} \right) \\ 
& ={{\left[ 1,{{e}^{j2\pi \left( \frac{{{d}_{t}}\sin {{\theta }_{t}}}{\lambda }-\frac{2\Delta f{{r}_{t}}}{c} \right)}},...,{{e}^{j2\pi \left( {{N}_{T}}-1 \right)\left( \frac{{{d}_{t}}\sin {{\theta }_{t}}}{\lambda }-\frac{2\Delta f{{r}_{t}}}{c} \right)}} \right]}^{T}} \\ 
\end{aligned}
\end{equation}
being the FDA range-angle-dependent transmit steering vector.
Let ${z_i}\left( t \right)$ denote the $i$-th undesired interference signal with $i = 1,2,...,\Upsilon $. For the $n$-th receiver element, the filtered output of the its ${m}'$-th channel may then be expressed as
\begin{equation}
{\bar z_{n,{m}';i}}\left( t \right) = {e^{ - j2\pi \frac{{\left( {n - 1} \right){d_r}\sin {\theta _i}}}{\lambda }}}{z_{{m}';i}}\left( t \right)
\end{equation}
for ${m}'=1,2,...,{{N}_{T}}$, where ${z_{{m}';i}}\left( t \right)$ is
\begin{equation}
{z_{m;i}}\left( t \right) = \left[ {{e^{ - j2\pi {f_m}t}}{z_i}\left( t \right)} \right] * {h_m}\left( t \right)
\end{equation}
with ${{h}_{{{m}'}}}\left( t \right)$ denoting the lowpass filter function.
Stacking all $N_T$ channel outputs yields
\begin{equation}
\begin{aligned}
{{{\mathbf{\bar{z}}}}_{n;i}}\left( t \right) &={{\left[ {{{\bar{z}}}_{n,1;i}}\left( t \right),{{{\bar{z}}}_{n,2;i}}\left( t \right),...,{{{\bar{z}}}_{n,{{N}_{T}};i}}\left( t \right) \right]}^{T}} \\ 
& ={{e}^{-j2\pi \frac{\left( n-1 \right){{d}_{r}}\sin {{\theta }_{i}}}{\lambda }}}{{\mathbf{z}}_{i}}\left( t \right) \\ 
\end{aligned}
\end{equation}
where ${{\mathbf{z}}_{i}}\left( t \right)=\left[ {{\mathbf{e}}^{c}}\left( t \right){{z}_{i}}\left( t \right) \right]*\mathbf{h}\left( t \right)$ with $^c$ and $*$ denoting the conjugate and convolution operators, respectively, and $\mathbf{h}\left( t \right)={{\left[ {{h}_{1}}\left( t \right),{{h}_{2}}\left( t \right),...,{{h}_{{{N}_{T}}}}\left( t \right) \right]}^{T}}$.
If introducing the presence of $\Upsilon$ undesired interferences and as well as an additive Gaussian noise, the FDA receive signal matrix may be obtained as 
\begin{subequations}\label{eq:1}
	\begin{equation}\label{eq:2}
\begin{aligned}
& {{\mathbf{R}}_{\text{FDA}}}\left( t \right) \\ 
& =\varsigma \left( {{r}_{t}},{{\theta }_{t}} \right)\operatorname{diag}\left\{ \mathbf{w}\odot {{\mathbf{a}}_{T}}\left( {{r}_{t}},{{\theta }_{t}} \right) \right\}\mathbf{s}\left( t-\frac{2{{r}_{t}}}{c} \right)\mathbf{b}_{R}^{T}\left( {{\theta }_{t}} \right) \\ 
& \kern 12pt +\sum\limits_{i=1}^{\Upsilon }{\left\{ {{\mathbf{z}}_{i}}\left( t \right)\mathbf{b}_{R}^{T}\left( {{\theta }_{i}} \right) \right\}}+\mathbf{N}\left( t \right) \\ 
\end{aligned} 
	\end{equation}
	\begin{equation}\label{eq:3}
\begin{aligned}
& =\varsigma \left( {{r}_{t}},{{\theta }_{t}} \right)\operatorname{diag}\left\{ \mathbf{s}\left( t-\frac{2{{r}_{t}}}{c} \right)\odot {{\mathbf{a}}_{T}}\left( {{r}_{t}},{{\theta }_{t}} \right) \right\}\mathbf{wb}_{R}^{T}\left( {{\theta }_{t}} \right) \\ 
& \kern 12pt +\sum\limits_{i=1}^{\Upsilon }{\left\{ {{\mathbf{z}}_{i}}\left( t \right)\mathbf{b}_{R}^{T}\left( {{\theta }_{i}} \right) \right\}}+\mathbf{N}\left( t \right) \\ 
\end{aligned} 
	\end{equation}
\end{subequations}
where $\mathbf{N}\left( t \right)$ represents the noise matrix, assuming it is both spatially and temporally white.
These two equivalent expressions for ${{\mathbf{R}}_{\text{FDA}}}\left( t \right)$
are used throughout the paper, i.e., the transmit waveform expression shown in \eqref{eq:2} and the transmit weight expression shown in \eqref{eq:3}.
For convenience, we stack the columns of ${{\mathbf{R}}_{\text{FDA}}}\left( t \right)$ to yield the vector 
\begin{align}
&{{{\mathbf{\bar{r}}}}_{\text{FDA}}}\left( t \right) \nonumber \\
&=\operatorname{vec}\left\{ {{\mathbf{R}}_{\text{FDA}}}\left( t \right) \right\} \nonumber \\
&=\varsigma \left( {{r}_{t}},{{\theta }_{t}} \right)\left[ {{\mathbf{b}}_{R}}\left( {{\theta }_{t}} \right)\otimes \operatorname{diag}\left\{ \mathbf{w}\odot {{\mathbf{a}}_{T}}\left( {{r}_{t}},{{\theta }_{t}} \right) \right\} \right]\mathbf{s}\left( t-\frac{2{{r}_{t}}}{c} \right) \nonumber \\
&\kern 12pt +\sum\limits_{i=1}^{\Upsilon }{\left\{ {{\mathbf{b}}_{R}}\left( {{\theta }_{t}} \right)\otimes {{\mathbf{z}}_{i}}\left( t \right) \right\}}+\mathbf{\bar{n}}\left( t \right) 
\end{align}
where $\mathbf{\bar{n}}\left( t \right)=\operatorname{vec}\left\{ \mathbf{N}\left( t \right) \right\}$, $\operatorname{vec}\left\{ \cdot  \right\}$ and $\otimes$ denote the vectorization operator and the Kronecker product, respectively.

\subsection{Interference Covariance Matrix and Interference Mitigation}

For the interference signal ${{z}_{{m}';i}}\left( t \right)$, according to Parseval's theorem,
\begin{equation}\label{eq:4}
\begin{aligned}
{{z}_{{m}';i}}\left( t \right) & =\left[ {{e}^{-j2\pi {{f}_{{m}'}}t}}{{z}_{i}}\left( t \right) \right]*{{h}_{{m}'}}\left( t \right) \\ 
& =\int\limits_{-\infty }^{\infty }{{{Z}_{i}}\left( f+{{f}_{{m}'}} \right){{H}_{{m}'}}\left( f \right){{e}^{j2\pi ft}}\text{d}f} \\ 
\end{aligned}
\end{equation}
where ${{H}_{{m}'}}\left( f \right)$ and ${{Z}_{i}}\left( f \right)$ are the Fourier transforms of ${{h}_{{m}'}}\left( t \right)$ and ${{z}_{i}}\left( t \right)$, respectively. Thus,
\begin{equation}
\begin{aligned}
\kern 6pt & \mathsf{\mathbb{E}}\left\{ {{z}_{{m}';i}}\left( t \right)z_{{m}'';i}^{c}\left( t \right) \right\} \\ 
& =\mathsf{\mathbb{E}}\left\{ \begin{aligned}
& \int\limits_{-\infty }^{\infty }{{{Z}_{i}}\left( f+{{f}_{{m}'}} \right){{H}_{{m}'}}\left( f \right){{e}^{j2\pi ft}}\text{d}f} \\ 
& \cdot \int\limits_{-\infty }^{\infty }{Z_{i}^{c}\left( {f}'+{{f}_{{m}''}} \right)H_{{{m}''}}^{c}\left( {{f}'} \right){{e}^{-j2\pi t{f}'}}\text{d}{f}'} \\ 
\end{aligned} \right\} \\ 
& =\int\limits_{-\infty }^{\infty }{\int\limits_{-\infty }^{\infty }{\left\{ \begin{aligned}
		& {{R}_{{{Z}_{i}}}}\left( f,{f}'' \right) \\ 
		& \cdot {{H}_{{m}'}}\left( f-{{f}_{{{m}'}}} \right)H_{{{m}''}}^{c}\left( {f}'-{{f}_{{{m}''}}} \right) \\ 
		& \cdot {{e}^{j2\pi \left( f-{{f}_{{m}'}} \right)t}}\cdot {{e}^{-j2\pi \left( {f}'-{{f}_{{{m}''}}} \right)t}} \\ 
		\end{aligned} \right\}\text{d}f\text{d}{f}'}} \\ 
\end{aligned}
\end{equation}
where $\mathsf{\mathbb{E}}\left\{ \cdot  \right\}$ represents the statistical expectation. Without loss of generality, suppose that ${{z}_{i}}\left( t \right)$ may be well modeled as a stationary
circularly symmetric complex random process with zero mean.
Following the results in \cite{9266663},
\begin{equation}
{{R}_{{{Z}_{i}}}}\left( f,{f}' \right)=\mathsf{\mathbb{E}}\left\{ {{Z}_{i}}\left( f \right)Z_{i}^{c}\left( {{f}'} \right) \right\}={{P}_{{{Z}_{i}}}}\left( f \right)\delta \left( f-{f}' \right).
\end{equation}
Substituted into $\left( 16 \right)$, this yields 
\begin{equation}
\begin{aligned}
& \mathsf{\mathbb{E}}\left\{ {{z}_{{m}';i}}\left( t \right)z_{{m}'';i}^{c}\left( t \right) \right\} = \\ 
& {{e}^{j2\pi \left( {{f}_{{{m}''}}}-{{f}_{{m}'}} \right)t}}\int\limits_{-\infty }^{\infty }{{{P}_{{{Z}_{i}}}}\left( f \right){{H}_{{m}'}}\left( f-{{f}_{{{m}'}}} \right)H_{{{m}''}}^{c}\left( f-{{f}_{{{m}''}}} \right)\text{d}f} \\ 
\end{aligned}
\end{equation}
Since the multi-channel mixing and low-pass filtering-based FDA receiver only works when the FDA transmit waveforms have non-overlapping spectrum, i.e.,
\begin{equation}
{{f}_{{m}'}}-{{f}_{{{m}''}}}>\frac{{{B}_{{{h}_{{m}'}}\left( t \right)}}+{{B}_{{{h}_{{{m}''}}}\left( t \right)}}}{2},
\end{equation} 
it holds that
\begin{equation}
\mathsf{\mathbb{E}}\left\{ {{z}_{{m}';i}}\left( t \right)z_{{m}'';i}^{c}\left( t \right) \right\}=\left\{ \begin{matrix}
{{P}_{{m}'}},{m}'={m}''  \\
0,{m}'\ne {m}''  \\
\end{matrix} \right.
\end{equation}
where ${{P}_{{m}'}}=\int\limits_{-\infty }^{\infty }{{{P}_{{{Z}_{i}}}}\left( f \right){{\left| {{H}_{{m}'}}\left( f+{{f}_{{m}'}} \right) \right|}^{2}}\text{d}f}$, which is a reasonable estimate for the interference power
spectrum density (PSD) at $f_{{m}'}$ when $N_T$ is sufficiently large.
From the derived results, it can be seen that the interference signal processed by the FDA receiver can be regarded as a spectral interference signal whose frequency is located at $f_{{m}'}$. It is worth noting
that this statistical property is different from conventional PA and MIMO radar systems.
Finally, the interference covariance matrix of the interference output by the FDA receiver, $\mathbf{Q}$, can be obtained as
\begin{equation}
\begin{aligned}
\mathbf{Q}& =\mathsf{\mathbb{E}}\left\{ \begin{aligned}
& \left[ \sum\limits_{i=1}^{\Upsilon }{\left\{ {{\mathbf{b}}_{R}}\left( {{\theta }_{i}} \right)\otimes {{\mathbf{z}}_{i}}\left( t \right) \right\}} \right] \\ 
& \cdot {{\left[ \sum\limits_{i=1}^{\Upsilon }{\left\{ {{\mathbf{b}}_{R}}\left( {{\theta }_{i}} \right)\otimes {{\mathbf{z}}_{i}}\left( t \right) \right\}} \right]}^{H}} \\ 
\end{aligned} \right\} =\sum\limits_{i=1}^{\Upsilon }{\left\{ {{\mathbf{Q}}_{i}} \right\}} \\ 
\end{aligned}
\end{equation}
where $^H$ is the conjugate transpose operator, and 
\begin{equation}
{{\mathbf{Q}}_{i}}={{\mathbf{P}}_{i}}\otimes \left[ {{\mathbf{b}}_{R}}\left( {{\theta }_{i}} \right)\mathbf{b}_{R}^{H}\left( {{\theta }_{t}} \right) \right],
\end{equation}
with ${{\mathbf{P}}_{i}}=\mathsf{\mathbb{E}}\left\{ {{\mathbf{z}}_{i}}\left( t \right)\mathbf{z}_{i}^{H}\left( t \right) \right\}$. Futhermore,
\begin{equation}
{{\mathbf{P}}_{i}}\left( l,k \right)=\left\{ \begin{matrix}
{{P}_{k}},l=k  \\
0,otherwise  \\
\end{matrix} \right.,
\end{equation}
resulting in 
${{\mathbf{P}}_{i}}=\sum\limits_{k=1}^{{{N}_{T}}}{{{P}_{k}} {{\mathbf{e}}_{k}}\mathbf{e}_{k}^{T}}$, where ${{\mathbf{e}}_{k}}$ indicates a vector with the $k$-th element being $1$ and $0$ otherwise.
Considering \eqref{eq:4}, the FDA snapshot ${{{\mathbf{\bar{r}}}}_\text{FDA}}\left( l \right)$ after analog-to-digital conversion (ADC) can therefore be expressed as
\begin{align}
&{{{\mathbf{\bar{r}}}}_{\text{FDA}}}\left( l \right) \nonumber=\varsigma \left( {{r}_{t}},{{\theta }_{t}} \right)\left[ {{\mathbf{b}}_{R}}\left( {{\theta }_{t}} \right)\otimes \operatorname{diag}\left\{ \mathbf{w}\odot {{\mathbf{a}}_{T}}\left( {{r}_{t}},{{\theta }_{t}} \right) \right\} \right]\mathbf{s}\left( l \right) \nonumber \\
&\kern 42pt +\sum\limits_{i=1}^{\Upsilon }{\left\{ {{\mathbf{b}}_{R}}\left( {{\theta }_{i}} \right)\otimes {{\mathbf{z}}_{i}}\left( l \right) \right\}}+\mathbf{\bar{n}}\left( l \right) \nonumber
\end{align}
with $l=1,2,...,L$, where $L$ denotes the number of discrete time samples.
Further, stacking all $L$ outputs yields
\begin{equation}
\begin{aligned}
 {{{\mathbf{\hat{r}}}}_{\text{FDA}}}&=\operatorname{vec}\left\{ \left[ {{{\mathbf{\bar{r}}}}_{\text{FDA}}}\left( l \right),{{{\mathbf{\bar{r}}}}_{\text{FDA}}}\left( 2 \right),...,{{{\mathbf{\bar{r}}}}_{\text{FDA}}}\left( L \right) \right] \right\} \\ 
& =\varsigma \left( {{r}_{t}},{{\theta }_{t}} \right)\mathbf{A}\left( {{r}_{t}},{{\theta }_{t}};\mathbf{w} \right)\mathbf{s}  +\sum\limits_{i=1}^{\Upsilon }{\left\{ \operatorname{vec}\left\{ \mathbf{I}\left( {{\theta }_{i}} \right) \right\} \right\}}+\mathbf{n}\\
& =\varsigma \left( {{r}_{t}},{{\theta }_{t}} \right)\mathbf{\tilde{A}}\left( {{r}_{t}},{{\theta }_{t}};\mathbf{s} \right)\mathbf{w}
+\sum\limits_{i=1}^{\Upsilon }{\left\{ \operatorname{vec}\left\{ \mathbf{I}\left( {{\theta }_{i}} \right) \right\} \right\}}+\mathbf{n} \\ 
\end{aligned}
\end{equation}
where
\begin{subequations}
	\begin{equation}
\kern 28pt \mathbf{I}\left( {{\theta }_{i}} \right)=\left[ {{\mathbf{b}}_{R}}\left( {{\theta }_{i}} \right)\otimes {{\mathbf{z}}_{i}}\left( 1 \right),...,{{\mathbf{b}}_{R}}\left( {{\theta }_{i}} \right)\otimes {{\mathbf{z}}_{i}}\left( L \right) \right]
	\end{equation}
	\begin{equation}
\begin{aligned}
& \kern -10pt \mathbf{A}\left( {{r}_{t}},{{\theta }_{t}};\mathbf{w} \right)={{\mathbf{I}}_{L}}\otimes \left[ {{\mathbf{b}}_{R}}\left( {{\theta }_{t}} \right)\otimes \operatorname{diag}\left\{ \mathbf{w}\odot {{\mathbf{a}}_{T}}\left( {{r}_{t}},{{\theta }_{t}} \right) \right\} \right] \\ 
\end{aligned}
	\end{equation}
	\begin{equation}
\begin{aligned}
 \kern -93pt \mathbf{\tilde{A}}\left( {{r}_{t}},{{\theta }_{t}};\mathbf{s} \right) =\left[ \begin{matrix}
\mathbf{\tilde{A}}\left( {{r}_{t}},{{\theta }_{t}};\mathbf{s}\left( l \right) \right)  \\
\mathbf{\tilde{A}}\left( {{r}_{t}},{{\theta }_{t}};\mathbf{s}\left( 2 \right) \right)  \\
...  \\
\mathbf{\tilde{A}}\left( {{r}_{t}},{{\theta }_{t}};\mathbf{s}\left( L \right) \right)  \\
\end{matrix} \right] \\ 
\end{aligned}
	\end{equation}
	\begin{equation}
\kern -8pt \mathbf{\tilde{A}}\left( {{r}_{t}},{{\theta }_{t}};\mathbf{s}\left( l \right) \right)=\left[ {{\mathbf{b}}_{R}}\left( {{\theta }_{t}} \right)\otimes \operatorname{diag}\left\{ \mathbf{s}\left( l \right)\odot {{\mathbf{a}}_{T}}\left( {{r}_{t}},{{\theta }_{t}} \right) \right\} \right]
	\end{equation}
		\begin{equation}
\kern -7pt \mathbf{n}=\operatorname{vec}\left\{ \left[ \mathbf{\bar{n}}\left( 1 \right),\mathbf{\bar{n}}\left( 1 \right),...,\mathbf{\bar{n}}\left( 1 \right) \right] \right\}
	\end{equation}
		\begin{equation}
\kern -9pt \mathbf{s}=\operatorname{vec}\left\{ \left[ \mathbf{s}\left( 1 \right),\mathbf{s}\left( 2 \right),...,\mathbf{s}\left( L \right) \right] \right\}
	\end{equation}
\end{subequations}
where $\mathbf{A}\left( {{r}_{t}},{{\theta }_{t}};\mathbf{w} \right) \in {{\mathbb{C}}^{{{N}_{T}}{{N}_{R}}L\times {{N}_{T}}L}}$ and $\mathbf{\tilde{A}}\left( {{r}_{t}},{{\theta }_{t}};\mathbf{s} \right) \in {{\mathbb{C}}^{{{N}_{T}}{{N}_{R}}L\times {{N}_{T}}}}$, with ${{\mathbb{C}}^{N\times N}}$ being the set of $N\times N$ complex-valued matrices.
Employing a receive weight vector $\mathbf{v}\in {{\mathbb{C}}^{{{N}_{T}}{{N}_{R}}L\times 1}}$ to synthesize the multi-channel outputs yields
\begin{align}
 {{\mathbf{x}}_{\text{FDA}}}&=\varsigma \left( {{r}_{t}},{{\theta }_{t}} \right){{\mathbf{v}}^{H}}\mathbf{A}\left( {{r}_{t}},{{\theta }_{t}};\mathbf{w} \right)\mathbf{s} \nonumber \\
&\kern 12pt  +\sum\limits_{i=1}^{\Upsilon }{\left\{ {{\mathbf{v}}^{H}}\operatorname{vec}\left\{ \mathbf{I}\left( {{\theta }_{i}} \right) \right\} \right\}}+{{\mathbf{v}}^{H}}\mathbf{n} \nonumber \\
& =\varsigma \left( {{r}_{t}},{{\theta }_{t}} \right){{\mathbf{v}}^{H}}\mathbf{\tilde{A}}\left( {{r}_{t}},{{\theta }_{t}};\mathbf{s} \right)\mathbf{w} \nonumber \\
&\kern 12pt +\sum\limits_{i=1}^{\Upsilon }{\left\{ {{\mathbf{v}}^{H}}\operatorname{vec}\left\{ \mathbf{I}\left( {{\theta }_{i}} \right) \right\} \right\}}+{{\mathbf{v}}^{H}}\mathbf{n}.
\end{align}
Consequently, the output SINR ${{\Pi }_{\text{FDA}}}\left( \mathbf{s},\mathbf{w},\mathbf{v} \right)$ can be written as
\begin{equation}
\begin{aligned}
& {{\Pi }_{\text{FDA}}}\left( \mathbf{s},\mathbf{w},\mathbf{v} \right) \\ 
& =\frac{\mathsf{\mathbb{E}}\left\{ {{\left| \varsigma \left( {{r}_{t}},{{\theta }_{t}} \right){{\mathbf{v}}^{H}}\mathbf{A}\left( {{r}_{t}},{{\theta }_{t}};\mathbf{w} \right)\mathbf{s} \right|}^{2}} \right\}}{\mathsf{\mathbb{E}}\left\{ \sum\limits_{i=1}^{\Upsilon }{\left\{ {{\mathbf{v}}^{H}}\operatorname{vec}\left\{ \mathbf{I}\left( {{\theta }_{i}} \right) \right\} \right\}} \right\}+\mathsf{\mathbb{E}}\left\{ {{\left| {{\mathbf{v}}^{H}}\mathbf{n} \right|}^{2}} \right\}} \\ 
& =\frac{\text{SNR}\cdot {{\left| {{\mathbf{v}}^{H}}\mathbf{A}\left( {{r}_{t}},{{\theta }_{t}};\mathbf{w} \right)\mathbf{s} \right|}^{2}}}{{{\mathbf{v}}^{H}}{{\mathbf{Q}}_{i+n}}\mathbf{v}}=\frac{\text{SNR}\cdot {{\left| {{\mathbf{v}}^{H}}\mathbf{\tilde{A}}\left( {{r}_{t}},{{\theta }_{t}};\mathbf{s} \right)\mathbf{w} \right|}^{2}}}{{{\mathbf{v}}^{H}}{{\mathbf{Q}}_{i+n}}\mathbf{v}} \\ 
\end{aligned}
\end{equation}
where
\begin{align}
{{{\mathbf{\bar{Q}}}}_{i+n}}&=\frac{1}{{{\sigma }^{2}}}\sum\limits_{i=1}^{\Upsilon }{\left\{ \mathsf{\mathbb{E}}\left\{ \left( \operatorname{vec}\left\{ \mathbf{I}\left( {{\theta }_{i}} \right) \right\} \right){{\left( \operatorname{vec}\left\{ \mathbf{I}\left( {{\theta }_{i}} \right) \right\} \right)}^{H}} \right\} \right\}}  \nonumber \\
&\kern 12pt +{{\mathbf{I}}_{{{N}_{T}}{{N}_{R}}L}} \nonumber \\
& =\frac{1}{{{\sigma }^{2}}}\sum\limits_{i=1}^{\Upsilon }{\left\{ \left( {{\mathbf{1}}_{L\times 1}}\mathbf{1}_{L\times 1}^{T} \right)\otimes {{\mathbf{Q}}_{i}} \right\}}+{{\mathbf{I}}_{{{N}_{T}}{{N}_{R}}L}} 
\end{align}
and where $\text{SNR=}\frac{\mathsf{\mathbb{E}}\left\{ {{\left| \varsigma \left( {{r}_{t}},{{\theta }_{t}} \right) \right|}^{2}} \right\}}{{{\sigma }^{2}}}$, with ${{\sigma }^{2}}$ denoting the noise power.
To suppress the present interferences, the adaptive minimum variance distortionless response (MVDR) receive weight vector ${{\mathbf{v}}_{opt}}$ is  employed, with
\begin{align}
{{\mathbf{v}}_{opt}}&=\frac{{{{\mathbf{\bar{Q}}}}_{i+n}}^{-1}\mathbf{A}\left( {{r}_{t}},{{\theta }_{t}};\mathbf{w} \right)\mathbf{s}}{{{\mathbf{s}}^{H}}{{\mathbf{A}}^{H}}\left( {{r}_{t}},{{\theta }_{t}};\mathbf{w} \right)\mathbf{Q}_{i+n}^{-1}\mathbf{A}\left( {{r}_{t}},{{\theta }_{t}};\mathbf{w} \right)\mathbf{s}} \nonumber \\
& =\frac{{{{\mathbf{\bar{Q}}}}_{i+n}}^{-1}\mathbf{\tilde{A}}\left( {{r}_{t}},{{\theta }_{t}};\mathbf{s} \right)\mathbf{w}}{{{\mathbf{w}}^{H}}{{{\mathbf{\tilde{A}}}}^{H}}\left( {{r}_{t}},{{\theta }_{t}};\mathbf{s} \right)\mathbf{Q}_{i+n}^{-1}\mathbf{\tilde{A}}\left( {{r}_{t}},{{\theta }_{t}};\mathbf{s} \right)\mathbf{w}} 
\end{align}
The corresponding output SINR may then be expressed as
\begin{equation}
{{\Pi }_{\text{FDA}}}\left( \mathbf{s},\mathbf{w} \right)={{\mathbf{s}}^{H}}\mathbf{\Psi }\left( \mathbf{w} \right)\mathbf{s}={{\mathbf{w}}^{H}}\mathbf{\tilde{\Psi }}\left( \mathbf{s} \right)\mathbf{w}
\end{equation}
where
\begin{subequations}\label{eq.7}
	\begin{equation}\label{eq.71}
\kern 3pt \mathbf{\Psi }\left( \mathbf{w} \right)={{\mathbf{A}}^{H}}\left( {{r}_{t}},{{\theta }_{t}};\mathbf{w} \right)\mathbf{Q}_{i+n}^{-1}\mathbf{A}\left( {{r}_{t}},{{\theta }_{t}};\mathbf{w} \right),
	\end{equation}
	\begin{equation}\label{eq.72}
\mathbf{\tilde{\Psi }}\left( \mathbf{s} \right)={{{\mathbf{\tilde{A}}}}^{H}}\left( {{r}_{t}},{{\theta }_{t}};\mathbf{s} \right)\mathbf{Q}_{i+n}^{-1}\mathbf{\tilde{A}}\left( {{r}_{t}},{{\theta }_{t}};\mathbf{s} \right).
	\end{equation}
\end{subequations}
It is worth noting that the output SINR is a function of the transmit waveforms and the weights.
We proceed to jointly designing the FDA transmit waveforms and weights with the aim of improving the output SINR, as well as limiting the waveform energy over the overlaid frequency bands.

\section{Problem Formulation}
\label{sec3}
In this section, we derive the FDA spectrum compatibility constraint and formulate the proposed method for constructing the transmit waveforms and weights accounting for various practical requirements. 

\subsection{Constraints}

\subsubsection{Energy and Bandwidth Constraints}
Both energy and bandwidth constraints must be imposed on the transmit waveforms to allow these to be separated by the receiver.
According to the results in \cite{1111111}, 
the energy constraint can be expressed as
\begin{equation}\label{eq.6}
\mathbf{s}_{T}^{H}{{\mathbf{\Sigma }}_{m}}{{\mathbf{s}}_{T}}=\frac{1}{{{N}_{T}}},
\end{equation}
for $m=1,2,...,{{N}_{T}}$, where ${{\mathbf{\Sigma }}_{m}}\in {{\mathbb{C}}^{{{N}_{T}}L\times {{N}_{T}}L}}$
represents a block diagonal matrix, with its $m$-th diagonal block being an identity matrix ${{\bf{I}}_{L}}$, and where ${{\mathbf{s}}_{T}}=\mathbf{T }\left( {{N}_{T}}L \right)\mathbf{s}$, with $\mathbf{T}\left( {{N}_{T}},L \right)=\sum\limits_{l=1}^{L}{\left( \mathbf{e}_{j}^{T}\otimes {{\mathbf{I}}_{{{N}_{T}}}}\otimes {{\mathbf{e}}_{j}} \right)}$ being the commutation matrix.
The bandwidth constraint has the form
\begin{equation}
{\bf{s}}_T^H{{\bf{B}}_m}{{\bf{s}}_T} \ge \frac{{{\gamma _m}}}{{{N_T}}},
\end{equation}
for $m = 1,2,...,{N_T}$,
where ${{\gamma }_{m}}\in \left( 0,1 \right]$ is a user-defined scalar that defines the
tolerance for in-band energy, a typical choice being ${{\gamma }_{m}}=0.91$ \cite{1111111}, whereas ${{\mathbf{B}}_{m}}$
represents a block diagonal matrix with its $m$-th diagonal block being
\begin{equation}
\int_{0}^{{{f}_{m,lp}}}{{{{\mathbf{\tilde{e}}}}_{f}}\mathbf{\tilde{e}}_{f}^{H}\text{d}f}\in {{\mathbb{C}}^{L\times L}}=\left\{ \begin{matrix}
{{f}_{m,lp}} & p=q  \\
\frac{{{e}^{j2\pi {{f}_{m,lp}}\left( p-q \right)}}-1}{j2\pi \left( p-q \right)} & p\ne q  \\
\end{matrix} \right.
\end{equation}
where ${{f}_{m,lp}}$ denotes the normalized cut-off frequency of $m$-th lowpass filter and ${{\bf{\tilde e}}_f} = {\left[ {{e^{ - j2\pi f}},{e^{ - j2\pi 2f}},...,{e^{ - j2\pi Lf}}} \right]^T}$.

\subsubsection{Similarity Constraint}
Implementing similarity constraint allows one to 
indirectly control desirable features of the transmit waveforms and weights.
For the transmit waveforms,
\begin{equation}
\left\| {{\mathbf{s}}_{T}}-{{\mathbf{s}}_{\operatorname{Ref}}} \right\|_{2}^{2}\le {{\varepsilon }^{2}},
\end{equation}
where $\left\| \cdot  \right\|_{2}^{2}$ represents the square of the Euclidean norm, ${{\mathbf{s}}_{\text{Ref}}}$ the reference waveform, and ${{\varepsilon }^{2}}$ a user-defined parameter to control the extend of the similarity.
The similarity constraint for the transmit weights is similarly
\begin{equation}
\left\| {{\mathbf{w}}_{T}}-{{\mathbf{w}}_{\operatorname{Ref}}} \right\|_{2}^{2}\le {{\mu }^{2}},
\end{equation}
where ${{\mathbf{w}}_{\text{Ref}}}$ and ${{\mu }^{2}}$ denote the reference weight and similarity level, respectively.

\subsubsection{Spectrally Compatible Constraint}
\begin{figure}[t]
	\centering
	\includegraphics[width=0.45\textwidth]{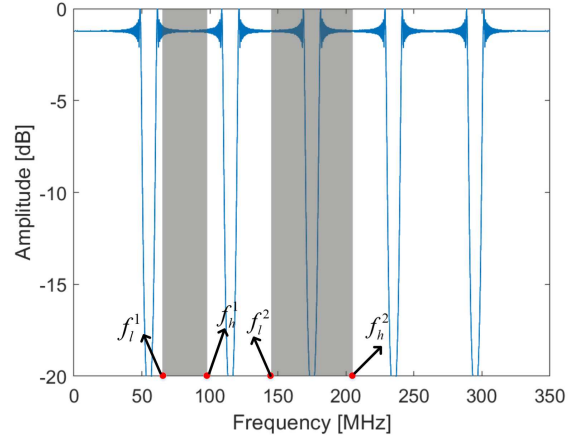}
	\caption{A diagram illustrating the FDA spectrum. Therein, the shared frequency bands are shaded in heavy gray, and $f_{l}^{1}$ and $f_{h}^{1}$ denote the lower and upper frequencies for the first shared system, respectively. $f_{l}^{2}$ and $f_{h}^{2}$ for the second shared system.}
	\label{Fi2}	
\end{figure}
Aside from the above requirements, the FDA needs to control the energy radiated on the shared frequency bands, allowing for more efficient use of the spectrum for the coexistence of different services.
Although it is a multi-waveform emission mechanism, the energy distribution of the FDA is actually direction-independent due to its non-overlapping waveform spectrum.
\begin{table}
	\centering
	\caption{FDA parameters.}
	{\begin{tabular}[l]{@{}ll}
			\toprule
			Parameter & Value\\
			\midrule
			Number of transmit antennas &$6$ \\
			Baseband Signal & LFM signal  \\
			Duration & $20 \kern 2pt \mu s$  \\
			Bandwidth of baseband Signal &$50 \kern 2pt MHz$ \\
			Frequency increment &$60 \kern 2pt MHz$  \\
			Total bandwidth &  $350 \kern 2pt MHz$\\
			Sampling frequency &$350 \kern 2pt MHz$\\
			Number of samples &$7000$\\
			\bottomrule
	\end{tabular}}
	\label{symbols}
\end{table}

An example of an FDA spectrum is shown in Fig. \ref{Fi2}, with the used parameters being listed in Tab. \ref{symbols}.
As can be seen the allocated frequency bands for each FDA transmit waveform are different, which would not be the case for a MIMO or a PA radar, indicating that the representation of the energy spectral density (ESD)-based spectral compatibility constraint should be adjusted \cite{7838312}.
Supposing that the $b$-th compatible electromagnetic system is operating over a frequency band ${{F}_{b}}=\left[ f_{l}^{b},f_{h}^{b} \right]$, let 
\begin{subequations}
	\begin{equation}
{{p}_{b}}=\left\lfloor \frac{f_{l}^{b}}{\Delta f} \right\rfloor,
	\end{equation}
	\begin{equation}
{{\tilde{p}}_{b}}=\left\lfloor \frac{f_{h}^{b}}{\Delta f} \right\rfloor
	\end{equation}
\end{subequations}
for $b=1,...,\Omega$, where $\left\lfloor \cdot  \right\rfloor $ is the rounding-down operator.
Thus, the FDA spectrally compatibly constraint can be expressed as (see Appendix)
\begin{equation}
{E_b} = {{\bf{w}}^H}{{\bf{\tilde I}}_b}\left( {\bf{S}} \right){\bf{w}} = {\bf{s}}_T^H{{\bf{H}}_b}\left( {\bf{w}} \right){{\bf{s}}_T} \le {\eta _b}
\end{equation}
where ${\eta _b}$ represents the acceptable level for the $b$-th compatible systems, and
${{\bf{\tilde I}}_b}\left( {\bf{S}} \right)$
and ${{\bf{H}}_b}\left( {\bf{w}} \right)$ are given in \eqref{eq:51} and \eqref{eq:55}, respectively.

\subsection{Optimization Problem}
Based on the aforementioned discussions, aiming to maximize the output SINR and to control the energy over the shared frequency bands, the joint design of transmit weights and waveforms for FDA can be formulated as
\begin{equation}
{{\mathrm{P}}_{1}}\left\{ \begin{matrix}
\underset{\mathbf{s}_T,\mathbf{w}}{\mathop{\max }}\, & \begin{aligned}
& {{\Pi }_\text{FDA}}\left( \mathbf{s},\mathbf{w} \right) \\ 
\end{aligned}  \\
s.t. & \left\{ \begin{matrix}
\left\| {{\mathbf{s}}_{T}}-{{\mathbf{s}}_{\operatorname{Ref}}} \right\|_{2}^{2}\le {{\varepsilon }^{2}}  \\
\mathbf{s}_{T}^{H}{{\mathbf{\Sigma }}_{m}}{{\mathbf{s}}_{T}}=\frac{1}{{{N}_{T}}}  \\
\mathbf{s}_{T}^{H}{{\mathbf{B}}_{m}}{{\mathbf{s}}_{T}}\ge \frac{{{\gamma }_{m}}}{{{N}_{T}}},{{\gamma }_{m}}\in \left( 0,1 \right]  \\
\left\| {{\mathbf{w}}}-{{\mathbf{w}}_{\operatorname{Ref}}} \right\|_{2}^{2}\le {{\mu }^{2}}  \\
\left\| {{\mathbf{w}}} \right\|_{2}^{2}={N_T}  \\
{{E}_{b}}={{\mathbf{w}}^{H}}{{{\mathbf{\tilde{I}}}}_{b}}\left( \mathbf{S} \right)\mathbf{w}=\mathbf{s}_{T}^{H}{{\mathbf{H}}_{b}}\left( \mathbf{w} \right){{\mathbf{s}}_{T}}\le {{\eta }_{b}}  \\
\end{matrix} \right.  \\
\end{matrix} \right.
\end{equation}
for $m=1,2,...,{{N}_{T}}$, where ${{\Pi }_\text{FDA}}\left( \mathbf{s},\mathbf{w} \right)$ denotes the output SINR given in \eqref{eq.6}.
It should be noted that optimizing the transmit waveforms with fixed transmit weights will achieve the FDA spectral compatibly design while still maximizing the output SINR.
However, as compared to the joint optimization, there are obvious disadvantages in controlling the transmission energy in the overlaid frequency bands and maintaining a constant energy in the transmit elements, which will also be demonstrated in the simulations below.

Generally speaking, there is no closed-form solution to $P_1$, as the energy and bandwidth constraints on the waveforms are nonconvex, as are the energy constraint on the weights and the objective function.
In the following, we will examine how $P_1$ may be reformulated to allow for an approximative solution using the proposed iterative algorithm.

\section{proposed algorithm}
\label{sec4}
In order to relax $P_1$, we separate the maximization, such that this is done alternatingly with respect to $\mathbf{w}$ and ${{\mathbf{s}}_{T}}$,
keeping the other fixed.
Each of the resulting problems may then be further relaxed.
In the following, we will examine both of these problems separately.

\subsection{Optimize $\mathbf{w}$ with fixed ${{\mathbf{s}}_{T}}$}

For a fixed transmit waveform vector ${{\mathbf{s}}_{T}}$, problem $P_1$ can be optimized with respect to $\mathbf{w}$ as
\begin{equation}
{{\mathrm{P}}_{2}}\left\{ \begin{matrix}
\underset{\mathbf{w}}{\mathop{\max }}\, & {{\mathbf{w}}^{H}}\mathbf{\tilde{\Psi }}\left( \mathbf{s} \right)\mathbf{w}  \\
s.t. & \left\{ \begin{matrix}
\left\| \mathbf{w} \right\|_{2}^{2}={N_T}  \\
\left\| \mathbf{w}-{{\mathbf{w}}_{\operatorname{Ref}}} \right\|_{2}^{2}\le {{\mu }^{2}}  \\
{{\mathbf{w}}^{H}}{{{\mathbf{\tilde{I}}}}_{b}}\left( \mathbf{S} \right)\mathbf{w}\le {{\eta }_{b}}  \\
\end{matrix} \right.  \\
\end{matrix} \right..
\end{equation}
Following the strategies in \cite{7414411,8356676}, this may be reformulated as
\begin{equation}
{{\mathrm{P}}_{3}}\left\{ \begin{matrix}
\underset{\mathbf{w}}{\mathop{\max }}\, & {{\mathbf{w}}^{H}}\mathbf{\tilde{\Psi }}\left( \mathbf{s} \right)\mathbf{w}  \\
s.t. & \left\{ \begin{matrix}
\left\| {{\mathbf{w}}_{T}} \right\|_{2}^{2}={N_T} \\
{{\mathbf{w}}^{H}}\left( {{\mathbf{I}}_{{{N}_{T}}}}-{{\mathbf{w}}_{\operatorname{Ref}}}\mathbf{w}_{\operatorname{Ref}}^{H} \right)\mathbf{w}\le {{\mu }^{2}}  \\
{{\mathbf{w}}^{H}}{{{\mathbf{\tilde{I}}}}_{b}}\left( \mathbf{S} \right)\mathbf{w}\le {{\eta }_{b}}  \\
\end{matrix} \right.  \\
\end{matrix} \right..
\end{equation}
Introducing a new variable $\mathbf{W}=\mathbf{w}{{\mathbf{w}}^{H}}$ yields
\begin{equation}
{{\mathrm{P}}_{4}}\left\{ \begin{matrix}
\underset{\mathbf{W}}{\mathop{\max }}\, & \operatorname{Tr}\left\{ \mathbf{\tilde{\Psi }}\left( \mathbf{s} \right)\mathbf{W} \right\}  \\
s.t. & \left\{ \begin{matrix}
\operatorname{Tr}\left\{ \mathbf{W} \right\}={N_T}  \\
\operatorname{Tr}\left\{ \left( {{\mathbf{I}}_{{{N}_{T}}}}-{{\mathbf{w}}_{\operatorname{Ref}}}\mathbf{w}_{\operatorname{Ref}}^{H} \right)\mathbf{W} \right\}\le {{\mu }^{2}}  \\
\operatorname{Tr}\left\{ {{{\mathbf{\tilde{I}}}}_{b}}\left( \mathbf{S} \right)\mathbf{W} \right\}\le {{\eta }_{b}}  \\
\mathbf{W}\succeq 0,\operatorname{Rank}\left\{ \mathbf{W} \right\}=1  \\
\end{matrix} \right.  \\
\end{matrix} \right.
\end{equation}
where $\mathbf{W}\succeq 0$
means that $\mathbf{W}$ is positive semidefinite, and with $\operatorname{Rank}\left\{ \mathbf{W} \right\}$ and $\operatorname{Tr}\left\{ \mathbf{W}  \right\}$ denoting the rank and trace of $\mathbf{W}$, respectively.
Reformulating $P_2$ to $P_4$ shows that the only nonconvex part of the problem is the rank constraint.
A natural relaxation is thus obtained by dropping this constraint, yielding the semidefinite program (SDP)
\begin{equation}
{{\mathrm{P}}_{5}}\left\{ \begin{matrix}
\underset{\mathbf{W}}{\mathop{\max }}\, & \operatorname{Tr}\left\{ \mathbf{\tilde{\Psi }}\left( \mathbf{s} \right)\mathbf{W} \right\}  \\
s.t. & \left\{ \begin{matrix}
\operatorname{Tr}\left\{ \mathbf{W} \right\}={N_T}  \\
\operatorname{Tr}\left\{ \left( {{\mathbf{I}}_{{{N}_{T}}}}-{{\mathbf{w}}_{\operatorname{Ref}}}\mathbf{w}_{\operatorname{Ref}}^{H} \right)\mathbf{W} \right\}\le {{\mu }^{2}}  \\
\operatorname{Tr}\left\{ {{{\mathbf{\tilde{I}}}}_{b}}\left( \mathbf{S} \right)\mathbf{W} \right\}\le {{\eta }_{b}}  \\
\mathbf{W}\succeq 0  \\
\end{matrix} \right.  \\
\end{matrix} \right.
\end{equation}
As the resulting problem is convex, a solution may be obtained in polynomial time using standard optimization tools, such as the convex optimization toolbox CVX \cite{grant2014cvx}.
If the rank of the globally optimal solution ${{\mathbf{W}}^{*}}$ obtained for problem $P_5$ is one, then ${{\mathbf{W}}^{*}}={{\mathbf{w}}^{*}}{{\mathbf{w}}^{*}}^{H}$, with ${{\mathbf{w}}^{*}}$ being the globally optimal solution to problem $P_2$.
Otherwise, a fundamental issue that one must address is how to convert the solution ${{\mathbf{W}}^{*}}$ into a feasible solution for problem $P_2$. 
To this end, a randomization method will here be used, as described in Algorithm \ref{Al1}.
In essence, the key concept is to construct random vectors that share ${{\mathbf{W}}^{*}}$ as a covariance matrix. Each random vector is then rescaled appropriately to ensure feasibility.

\begin{algorithm}[t]
	\renewcommand{\algorithmicrequire}{\textbf{Input:}}
	\renewcommand{\algorithmicensure}{\textbf{Output:}}
	\caption{Algorithm to solve Problem $P_2$}
	\begin{algorithmic}
		\REQUIRE The overlaid frequency band ${{F}_{b}}$, the acceptable level ${{\eta }_{b}}$, $b=1,...,\Omega$, the similarity level ${{\mu }^{2}}$, the reference transmit weight vector ${{\mathbf{w}}_{\operatorname{Ref}}}$, and the initial transmit waveform matrix ${{\mathbf{S}}_{0}}$.\\ 
		\ENSURE A approximate solution ${{\mathbf{w}}^ * }$ of Problem $P_2$.
		\STATE $1$: Calculate ${{{\mathbf{\tilde{I}}}}_{b}}\left( {{\mathbf{S}}_{0}} \right)$ and $\mathbf{\tilde{\Psi }}\left( {{\mathbf{s}}_{0}} \right)$ using \eqref{eq:5} and \eqref{eq.72}.
		\STATE $2$: Solve the SDP problem $P_5$ yielding ${{\mathbf{W}}^{*}}$.  
		\STATE $3$: Perform the eigenvalue decomposition $ {{\mathbf{W}}^{*}}=\mathbf{V\Delta }{{\mathbf{V}}^{H}}$.
		\STATE $4$: If $\operatorname{Rank} \left\{ {\mathbf{V}} \right\} = 1$, then $\mathbf{V}={{\mathbf{w}}^{*}}{{\mathbf{w}}^{*}}^{H}$, Terminate.
		\STATE $5$: Otherwise, run the randomization method as follows:\\
		$\kern 10pt \romannumeral1)$ Perform a Cholesky factorization on ${{\mathbf{W}}^{*}}$, yielding \\
		$\kern 22pt {{\mathbf{W}}^{*}}=\mathbf{K}{{\mathbf{K}}^{H}}$\\
		$\kern 10pt \romannumeral2)$ Generate $J$ standard Gaussian random vectors ${{\boldsymbol{\xi }}_{j}}$, \\
		$\kern 24pt j=1,2...,J$ and set ${{\mathbf{p}}_{j}}={{\mathbf{K}}^{H}}{{\boldsymbol{\xi }}_{j}}$ \\
		$\kern 10pt \romannumeral3)$ Let ${{{\mathbf{\bar{p}}}}_{j}}=\frac{\sqrt{{{N}_{T}}}{{\mathbf{p}}_{j}}}{{{\left\| {{{\mathbf{\bar{p}}}}_{j}} \right\|}_{2}}}$ \\
		$\kern 10pt \romannumeral4)$ Extract elements ${{\mathbf{\tilde{p}}}_{q}},q=1,2...,,Q$, from set ${{D}_{\mathbf{w}}}$ \\
		$\kern 23pt $ defined as \\
		$\kern 35pt {{D}_{\mathbf{w}}}=\left\{ {{{\mathbf{\bar{p}}}}_{j}}\left| \begin{matrix}
		\mathbf{\bar{p}}_{j}^{H}\left( {{\mathbf{I}}_{{{N}_{T}}}}-{{\mathbf{w}}_\text{Ref}}\mathbf{w}_\text{Ref}^{H} \right){{{\mathbf{\bar{p}}}}_{j}}\le {{\mu }^{2}}  \\
		\mathbf{\bar{p}}_{j}^{H}{{{\mathbf{\tilde{I}}}}_{b}}\left( \mathbf{S} \right){{{\mathbf{\bar{p}}}}_{j}}\le {{\eta }_{b}}  \\
		\end{matrix} \right. \right\}$\\
		$\kern 23pt $ with $Q$ being the number of the feasible trials for \\
		$\kern 24pt$ the problem $P_2$\\ 
		$\kern 10pt \romannumeral5)$ Output \\
		$\kern 50pt {{\mathbf{w}}^{*}}=\arg \underset{{{{\mathbf{\tilde{p}}}}_{q}}}{\mathop{\max }}\,\mathbf{\tilde{p}}_{q}^{H}\mathbf{\tilde{\Psi }}\left( {{\mathbf{s}}_{0}} \right){{\mathbf{\tilde{p}}}_{q}}$
	\end{algorithmic}
\label{Al1}
\end{algorithm}

\subsection{Optimize ${\mathbf{s}_T}$ with fixed $\mathbf{w}$}
Proceeding, for a fixed transmit weight vector, $\mathbf{w}$, the optimal transmit waveform vector ${\mathbf{s}_T}$ can be
obtained by rewriting $P_1$ as
\begin{equation}
{{\mathrm{P}}_{6}}\left\{ \begin{matrix}
\underset{\mathbf{s}_T}{\mathop{\max }}\, & \mathbf{s}_{T}^{H}\mathbf{T }\left( {{N}_{T}},L \right)\mathbf{\Psi }\left( \mathbf{w} \right)\mathbf{T }\left( L,{{N}_{T}} \right){{\mathbf{s}}_{T}}  \\
s.t. & \left\{ \begin{matrix}
\left\| {{\mathbf{s}}_{T}}-{{\mathbf{s}}_{\operatorname{Ref}}} \right\|_{2}^{2}\le {{\varepsilon }^{2}}  \\
\mathbf{s}_{T}^{H}{{\mathbf{\Sigma }}_{m}}{{\mathbf{s}}_{T}}=\frac{1}{{{N}_{T}}}  \\
\mathbf{s}_{T}^{H}{{\mathbf{B}}_{m}}{{\mathbf{s}}_{T}}\ge \frac{{{\gamma }_{m}}}{{{N}_{T}}},{{\gamma }_{m}}\in \left( 0,1 \right]  \\
\mathbf{s}_{T}^{H}{{\mathbf{H}}_{b}}\left( \mathbf{w} \right){{\mathbf{s}}_{T}}\le {{\eta }_{b}}  \\
\end{matrix} \right.  \\
\end{matrix} \right.
\end{equation}
for $m=1,2,...,{{N}_{T}}$.
Let $\mathbf{\bar{S}}={{\mathbf{s}}_{T}}\mathbf{s}_{T}^{H}$, suggesting the relaxed SDP problem $P_6$ obtained by dropping the rank-one constraint, i.e.,
\begin{equation}
{{\mathrm{P}}_{7}}\left\{ \begin{matrix}
\underset{\mathbf{\bar{S}}}{\mathop{\max }}\, & \operatorname{Tr}\left\{ \mathbf{T }\left( {{N}_{T}},L \right)\mathbf{\Psi }\left( \mathbf{w} \right)\mathbf{T }\left( L,{{N}_{T}} \right)\mathbf{\bar{S}} \right\}  \\
s.t. & \left\{ \begin{matrix}
\operatorname{Tr}\left\{ {{\mathbf{\Sigma }}_{m}}\mathbf{\bar{S}} \right\}=\frac{1}{{{N}_{T}}}  \\
\operatorname{Tr}\left\{ \left( {{\mathbf{I}}_{{{N}_{T}}}}-{{\mathbf{s}}_{\operatorname{Ref}}}\mathbf{s}_{\text{Ref}}^{H} \right)\mathbf{\bar{S}} \right\}\le {{\varepsilon }^{2}}  \\
\operatorname{Tr}\left\{ {{\mathbf{B}}_{m}}\mathbf{\bar{S}} \right\}\ge \frac{{{\gamma }_{m}}}{{{N}_{T}}},{{\gamma }_{m}}\in \left( 0,1 \right]  \\
\operatorname{Tr}\left\{ {{\mathbf{H}}_{b}}\left( \mathbf{w} \right)\mathbf{\bar{S}} \right\}\le {{\eta }_{b}}  \\
\mathbf{\bar{S}}\succeq 0  \\
\end{matrix} \right.  \\
\end{matrix} \right.
\end{equation}
The resulting maximization is convex and the globally optimal solution  ${{\mathbf{\bar{S}}}^{*}}$ of problem $P_7$ may thus be obtained using standard optimization tools.
As above, a similar randomization technique may be employed to find an approximate solution to problem $P_6$.
The proposed iterative algorithm is summarized in Algorithm \ref{Al2}.

\begin{algorithm}[t]
	\renewcommand{\algorithmicrequire}{\textbf{Input:}}
	\renewcommand{\algorithmicensure}{\textbf{Output:}}
	\caption{SDR-based Algorithm for solving Problem $P_1$}
	\begin{algorithmic}
		\REQUIRE Set the in-band energy tolerance ${{\gamma }_{m}}$, the waveform similarity level ${{\varepsilon }^{2}}$, the reference waveforms ${{\mathbf{s}}_{\operatorname{Ref}}}$, and the number of iterations $G_{\max }$.\\ 
		\ENSURE The optimal transmit waveforms and weights pair\\
		 $ \kern 30pt \left( {{\mathbf{s}}_{T,opt}},{{\mathbf{w}}_{opt}} \right)$.\\
       Set $q=1$;\\
		 \WHILE{ $q\le {{G}_{\max }}$ }
		\STATE  $1$: Obtain ${{\mathbf{w}}^{q}}$ via \textbf{Algorithm 1};\\
		\STATE $2$: Substitute ${{\mathbf{w}}^{q}}$ into problem $P_6$ and solve $P_7$;\\
		\STATE $3$: Apply a randomization method to obtain an \\
		$ \kern 7.9pt$ approximate solution $\mathbf{s}_{T}^{q}$ of problem $P_6$;\\
		\STATE $4$: Let $\mathbf{s}_{T}^{q}{{\left( \mathbf{s}_{T}^{q} \right)}^{H}}$ be the initial transmit waveform matrix \\
		$ \kern 7.9pt$ of \textbf{Algorithm \ref{Al1}};\\
		\STATE $5$: Set $q=q+1$.
		\ENDWHILE
		\RETURN $\left( \mathbf{s}_{T}^{{{G}_{\max }}},{{\mathbf{w}}^{{{G}_{\max }}}} \right)=\left( {{\mathbf{s}}_{T,opt}},{{\mathbf{w}}_{opt}} \right)$.
	\end{algorithmic}
\label{Al2}
\end{algorithm}


\section{NUMERICAL SIMULATIONS}
\label{sec5}
In this section, various numerical simulations are provided to assess the performance of the proposed FDA design scheme.
We first consider the problem of the FDA spectrally compatible design with different design schemes, namely, designing only the transmit waveforms, designing only transmit weights, and joint design of the transmit waveforms and weights. We compare the output SINR and achieved ESD for these approaches over the overlaid frequency bands.
Next, we examine the ACF of the designed waveforms. Finally, the behavior of the FDA receive beampattern is illustrated. 
In all simulations, it is assumed that the target of interest is located at $\left( {{r}_{t}},{{\theta }_{t}} \right)=\left( 15 \kern 1pt km,{{40}^{\operatorname{o}}} \right)$ with the input SNR being $0 \kern2pt dB$.
Furthermore, the signal is assumed to be corrupted by these interference sources, whose details are listed in Tab. \ref{symbolss}.
We also assume $\Omega=2$ compliant systems operating on the normalized frequency bands ${{F}_{1}}=\left( 0.073,0.200 \right)$ and ${{F}_{2}}=\left( 0.556,0.884 \right)$ that overlap with FDA frequency band.
The assumed system parameters are listed in Tab. \ref{symbolsss}.
\begin{table}
	\centering
	\caption{Parameter setting for interferences}
	{\begin{tabular}[l]{@{}cccc}
			\toprule
			No. of interference & Frequency [MHz] & Angle [deg] & INR [dB]\\
			\midrule
			1 & 10000 & 10 & 20 \\
			2 & 10002 & 40 & 22 \\
			3 & 10004 & 60 & 24 \\
			\bottomrule
	\end{tabular}}
	\label{symbolss}
\end{table}

\begin{table}
	\centering
	\caption{System parameters}
	{\begin{tabular}[l]{@{}ll}
			\toprule
			Parameter & Value\\
			\midrule
			Number of transmit antennas &$N_T = 6$ \\
			Number of receive antennas &$N_R = 4$ \\
			Carrier frequency&$f_c = 10 \kern 2pt GHz$ \\
			Frequency increment & $\Delta f=1 \kern 2pt MHz$\\
			Duration & $40 \kern 2pt\mu s$\\
			Sampling frequency & $1 \kern 2pt MHz$\\
			Tolerance of in-band energy &  $\begin{aligned}
			& {{\gamma }_{m}}=0.91 \\ 
			& m=1,...,{{N}_{T}} \\ 
			\end{aligned}$\\
			\bottomrule
	\end{tabular}}
	\label{symbolsss}
\end{table}

\subsection{FDA Spectrally Compatible Design}
\subsubsection{Designing only the transmit waveforms}
\begin{figure}[t]
	\centering
	\includegraphics[width=0.46\textwidth, height=0.342\textwidth]{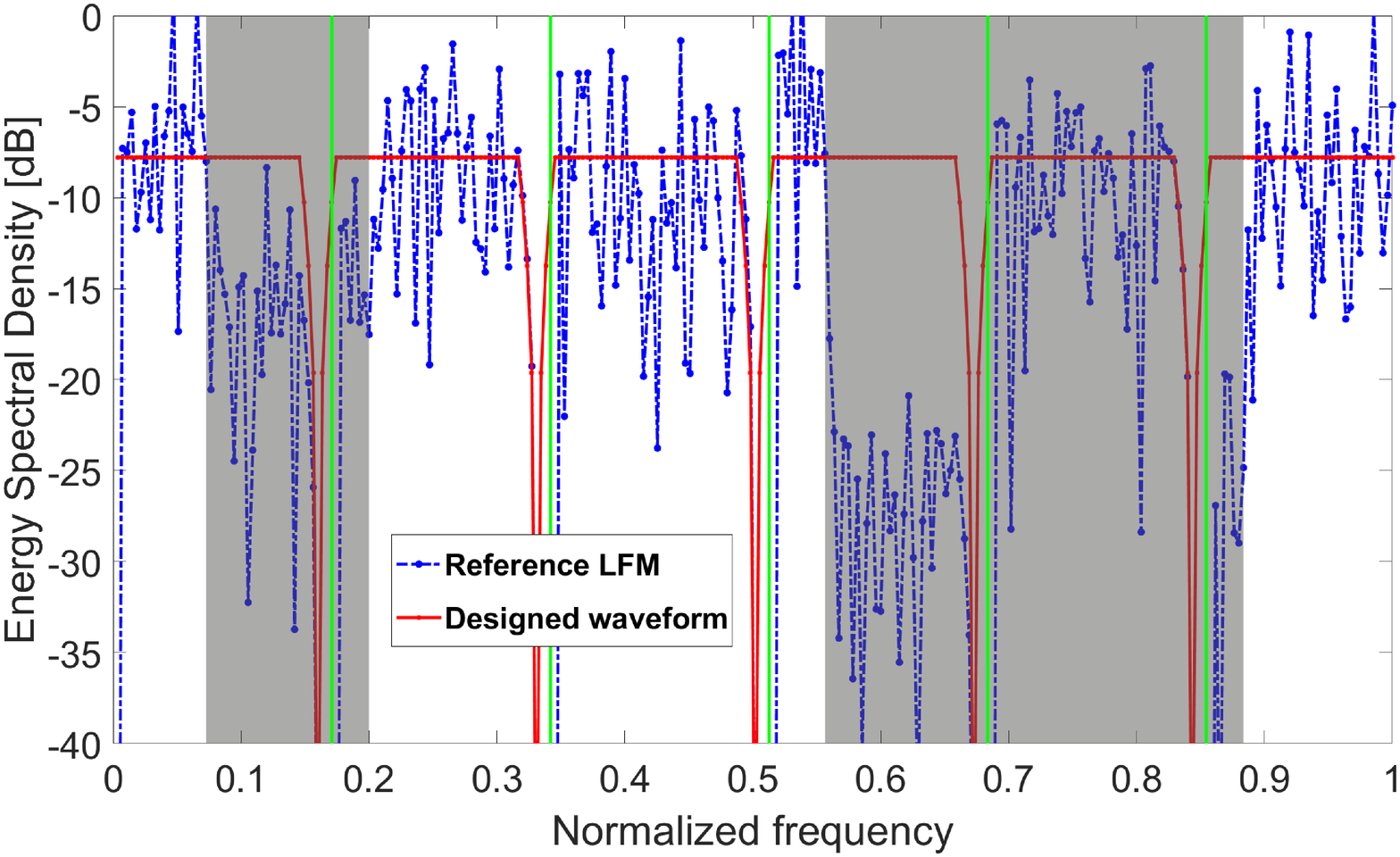}
	\caption{The ESD of the synthesized waveforms with a similarity value ${{\varepsilon }^{2}}=6$.}
	\label{Fi3}	
\end{figure}

Usually, the performance compatibility is evaluated by the amount of energy produced over the shared frequency bands \cite{6850145,7414411,7838312}.
For PA and MIMO systems, the compatibility requirements can be met by designing only the transmit waveform.
However, for FDA with its non-overlapping spectra, it is difficult to achieve the same requirement under a constant energy constraint.
In particular, for the second shared system, the normalized frequency band $F_2$ covers the spectrum of the transmit waveform $4$, $5$, and $6$, resulting in an achievable energy greater than $\frac{1}{{{N}_{T}}}$, which is not applicable in practice.
Here, we set the energy attenuation of the first frequency band $F_1$ to $10 \kern 2pt dB$, that is, ${{\eta }_{1}}=\frac{2}{{{N}_{T}}}\cdot \frac{1}{10}=\frac{1}{30}$.
At the same time, the allowable energy in the second band $F_2$ is set to ${{\eta }_{2}}=\frac{2}{{{N}_{T}}}\cdot \frac{1}{{{10}^{2}}}+\frac{1}{{{N}_{T}}}=\frac{101}{300}$.
Fig. \ref{Fi3} shows the ESD of the synthesized waveforms with a similarity value ${{\varepsilon }^{2}}=6$.
Therein, the shared frequency bands are shaded in heavy gray, and the green straight lines mark the frequency bands of the different transmit waveforms.
The result reveals that, for FDA, the scheme is capable of limiting the radiated energy on the shared bands, as required by the imposed spectrally compatible constraints.
But the achieved compatibility ${{\eta }_{2}}=\frac{101}{300}$ does not perform well on the second frequency band due to the constant energy limitation.

\subsubsection{Designing only the transmit weights}
\begin{figure}[t]
	\centering
	\includegraphics[width=0.45\textwidth]{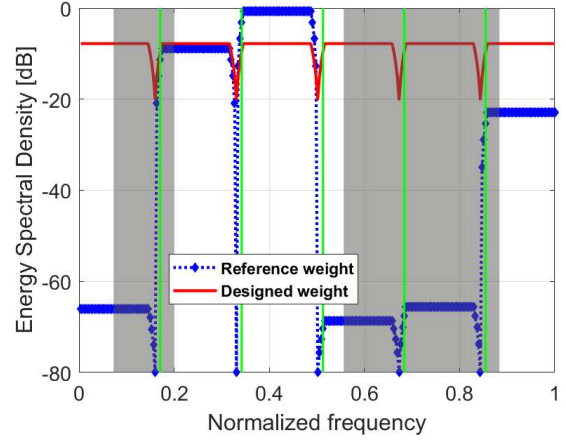}
	\caption{The obtained spectral distributions with similarity value ${{\mu }^{2}}=15$ when designing only the transmit weights.}
	\label{Fi4}
\end{figure}

\begin{figure}[t]
	\centering
	\includegraphics[width=0.45\textwidth]{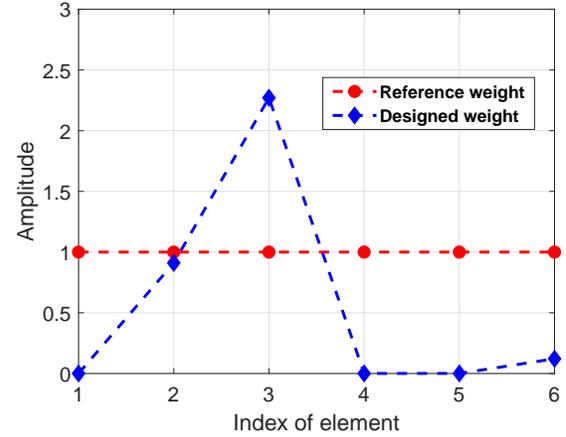}
	\caption{Comparison between the reference weights and the designed weights.}
	\label{Fi5}
\end{figure}

Proceeding, the energy attenuation on ${{F}_{1}}=\left( 0.073,0.200 \right)$ and ${{F}_{2}}=\left( 0.556,0.884 \right)$ are set to ${{\eta }_{1}}=\frac{2}{{{N}_{T}}}\cdot \frac{1}{10}=\frac{1}{30}$ and ${{\eta }_{2}}=\frac{3}{{{N}_{T}}}\cdot \frac{1}{{{10}^{2}}}=\frac{1}{200}$, respectively.
In Fig. \ref{Fi4}, the spectral distributions after designing the transmit weights with a similarity value ${{\mu }^{2}}=15$ is shown.
Benefiting from the non-overlapping spectrum, with flexible transmit power allocation, FDA can properly control the transmit power at the shared frequencies without adjusting the waveform.
Fig. \ref{Fi5} compares the designed weights and the reference weights. It can be seen that the designed weights are well matched to the frequency bands of the compatible systems. However, in practice, the severely non-constant energy exhibited will reduce the power efficiency of the amplifier and broaden the transmitted spectrum. At the same time, the SINR performance, as is further shown below, is greatly degraded.

\subsubsection{Jointly designing the transmit waveforms and weights}
\begin{figure*}[t]
	\centering
	\includegraphics[width=0.8\textwidth]{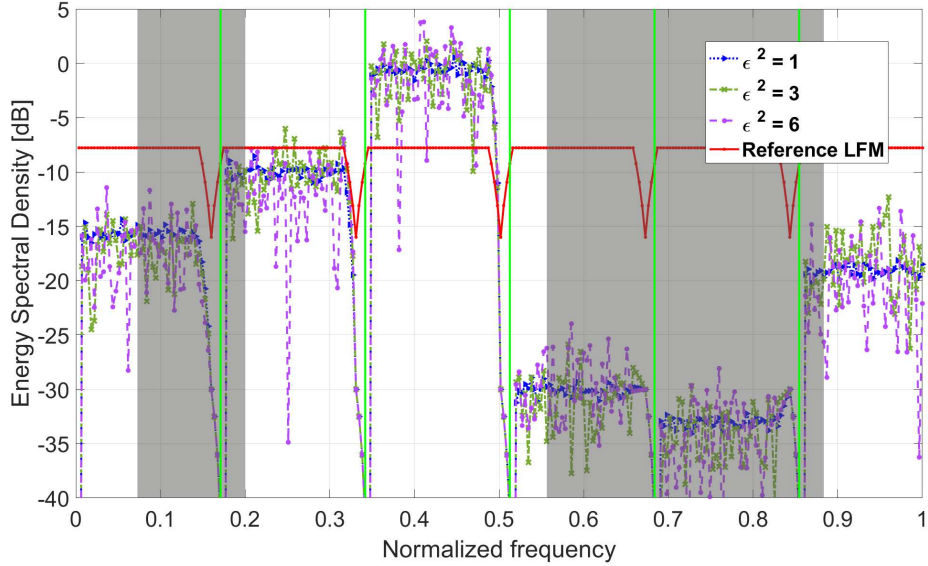}
	\caption{The obtained ESD for different transmit waveform similarity values with a fixed transmit weight similarity level ${{\mu }^{2}}=15$.}
	\label{Fi6}
\end{figure*}

\begin{figure}[t]
	\centering
	\includegraphics[width=0.45\textwidth]{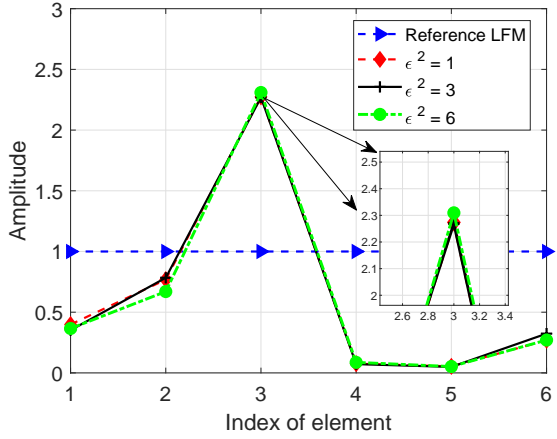}
	\caption{Comparison between the reference weights and the designed weights.}
	\label{Fi7}
\end{figure}

\begin{figure}[t]
	\centering
	\includegraphics[width=0.45\textwidth]{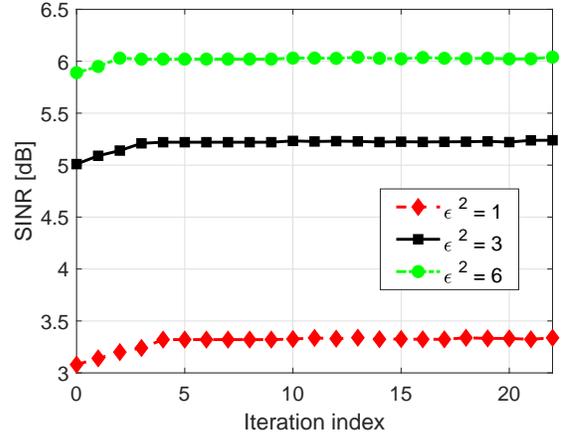}
	\caption{The output SINR versus the iteration number for different ${\varepsilon }^{2}$ with ${{\mu }^{2}}=15$.}
	\label{Fi8}
\end{figure}

With the same energy attenuation settings as in Fig. \ref{Fi4}, Fig. \ref{Fi6} illustrates the spectral behavior with a fixed transmit weight similarity level ${{\mu }^{2}}=15$ but with different transmit waveform similarity values ${{\varepsilon }^{2}}=1$, $3$, and $6$.
As expected, by jointly adjusting both the transmit waveforms and weights, a better energy distribution is achievable, as well as less variation in energy allocated to each transmit element, as shown in Fig. \ref{Fi7}.
Fig. \ref{Fi8} shows the output SINR for different transmit waveform similarity levels.
It is worth noting that the similarity value has a greater impact on the output SINR than the number of iterations.
This result may arise due to, for the proposed algorithm, the transmit weights and waveforms play similar roles, making it is difficult to improve the output SINR by increasing the number of iterations.

Tab. \ref{symbsols} summarizes the output SINRs for the different designs. 
It may be noted that the scheme that only optimizes the transmit waveforms achieves the worst spectral compatibility.
Meanwhile, the output SINR of the transmit weight design shows the worst performance due to the transmit degrees of freedom (DoF) of the FDA is being sacrificed in order to fulfill the spectral compatibility constraint.
Compared with the weight design, the joint design can obtain the same spectral compatibility, and compared with the waveform design, the joint design can obtain a slightly larger output SINR.
\begin{table*}
	\centering
	\caption{Performance comparison.}
	{\begin{tabular}[l]{@{}cccc}
			\toprule
			Type & Spectral compatibility & Output SINR [dB]\\
			\midrule
			Transmit wavefrom design &$\left\{ {{\eta }_{1}}=\frac{1}{30},{{\eta }_{2}}=\frac{101}{300} \right\}\leftrightarrow \left\{ 10 \kern 1.5pt dB,1.72 \kern 1.5pt dB \right\}$ & $5.99$ \\
			Transmit weight design &$\left\{ {{\eta }_{1}}=\frac{1}{30},{{\eta }_{2}}=\frac{1}{200} \right\}\leftrightarrow \left\{ 10 \kern 1.5pt dB,20 \kern 1.5pt dB \right\}$ & $2.36$ \\
			Joint design of transmit wavefroms and weights&$\left\{ {{\eta }_{1}}=\frac{1}{30},{{\eta }_{2}}=\frac{1}{200} \right\}\leftrightarrow \left\{ 10 \kern 1.5pt dB,20 \kern 1.5pt dB \right\}$ & $6.02$\\
			\bottomrule
	\end{tabular}}
	\label{symbsols}
\end{table*}

\subsection{ACF Property}
\begin{figure*}[t]
	\centering
	
	\subfigure[]{
		\begin{minipage}[t]{0.33\linewidth}
			\centering
			\includegraphics[width=0.22\textheight]{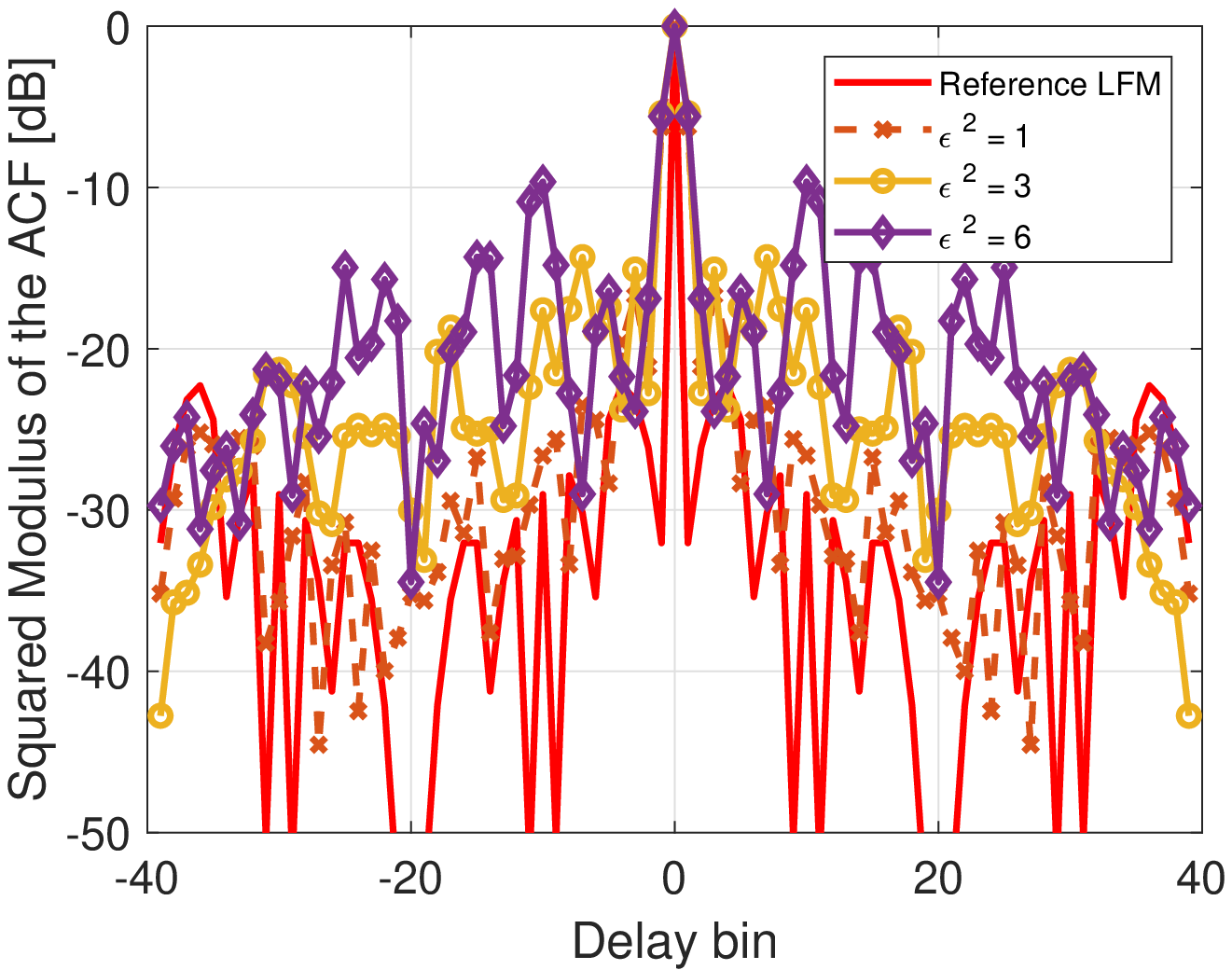}
		\end{minipage}%
	}%
	\subfigure[]{
		\begin{minipage}[t]{0.33\linewidth}
			\centering
			\includegraphics[width=0.22\textheight]{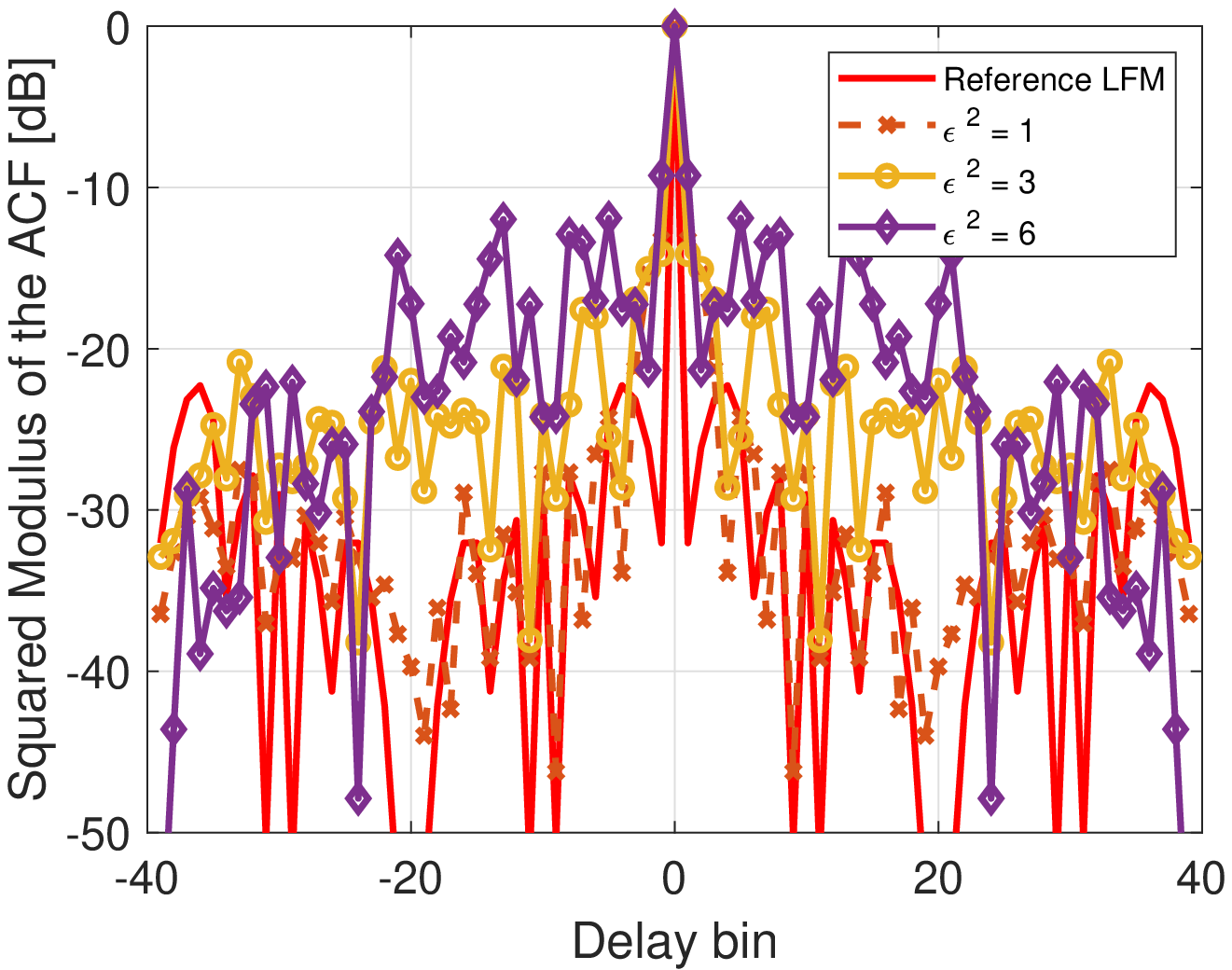}
		\end{minipage}%
	}%
	\subfigure[]{
		\begin{minipage}[t]{0.33\linewidth}
			\centering
			\includegraphics[width=0.22\textheight]{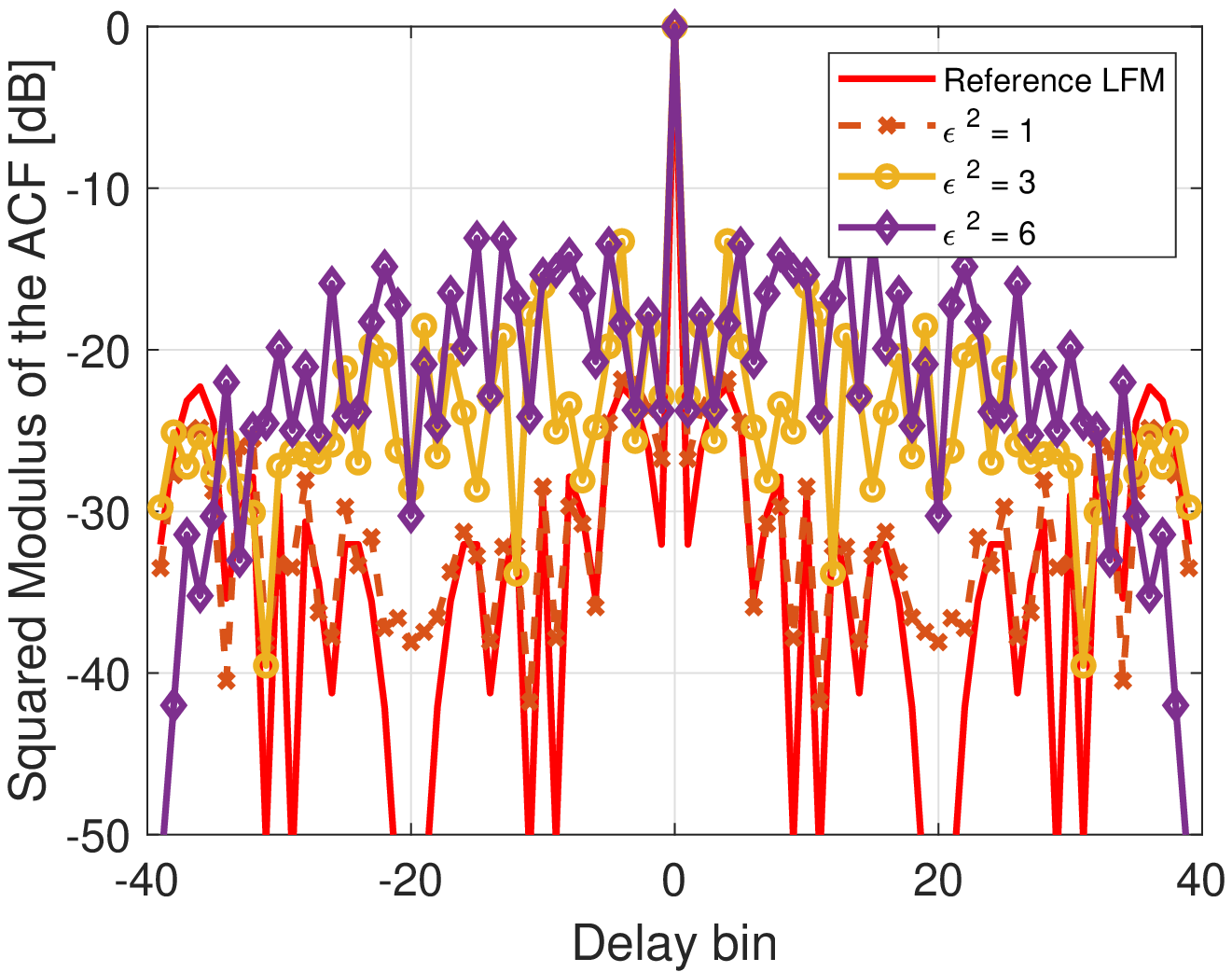}
		\end{minipage}%
	}%
	
	\subfigure[]{
		\begin{minipage}[t]{0.33\linewidth}
			\centering
			\includegraphics[width=0.22\textheight]{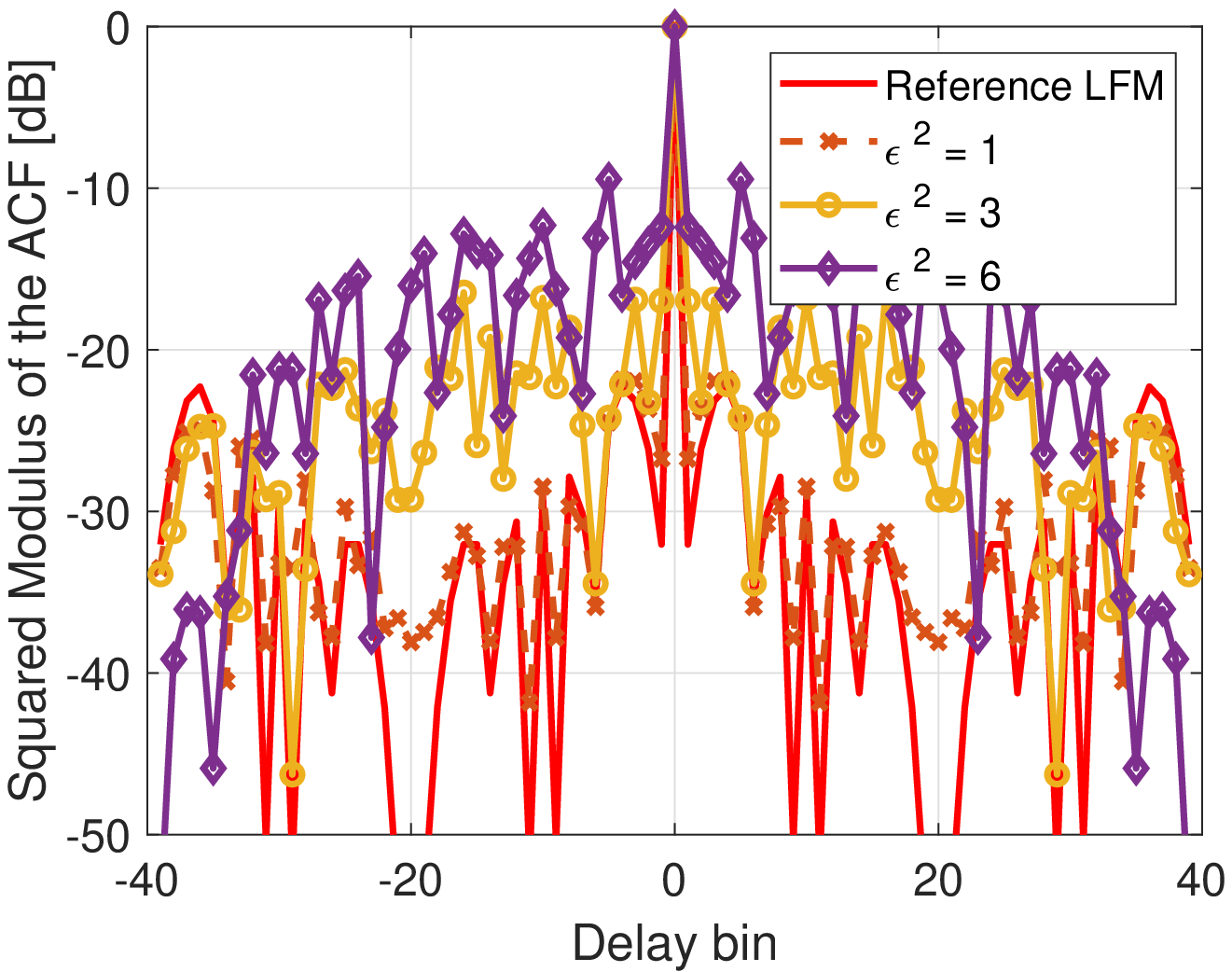}
		\end{minipage}
	}%
	\subfigure[]{
		\begin{minipage}[t]{0.33\linewidth}
			\centering
			\includegraphics[width=0.22\textheight]{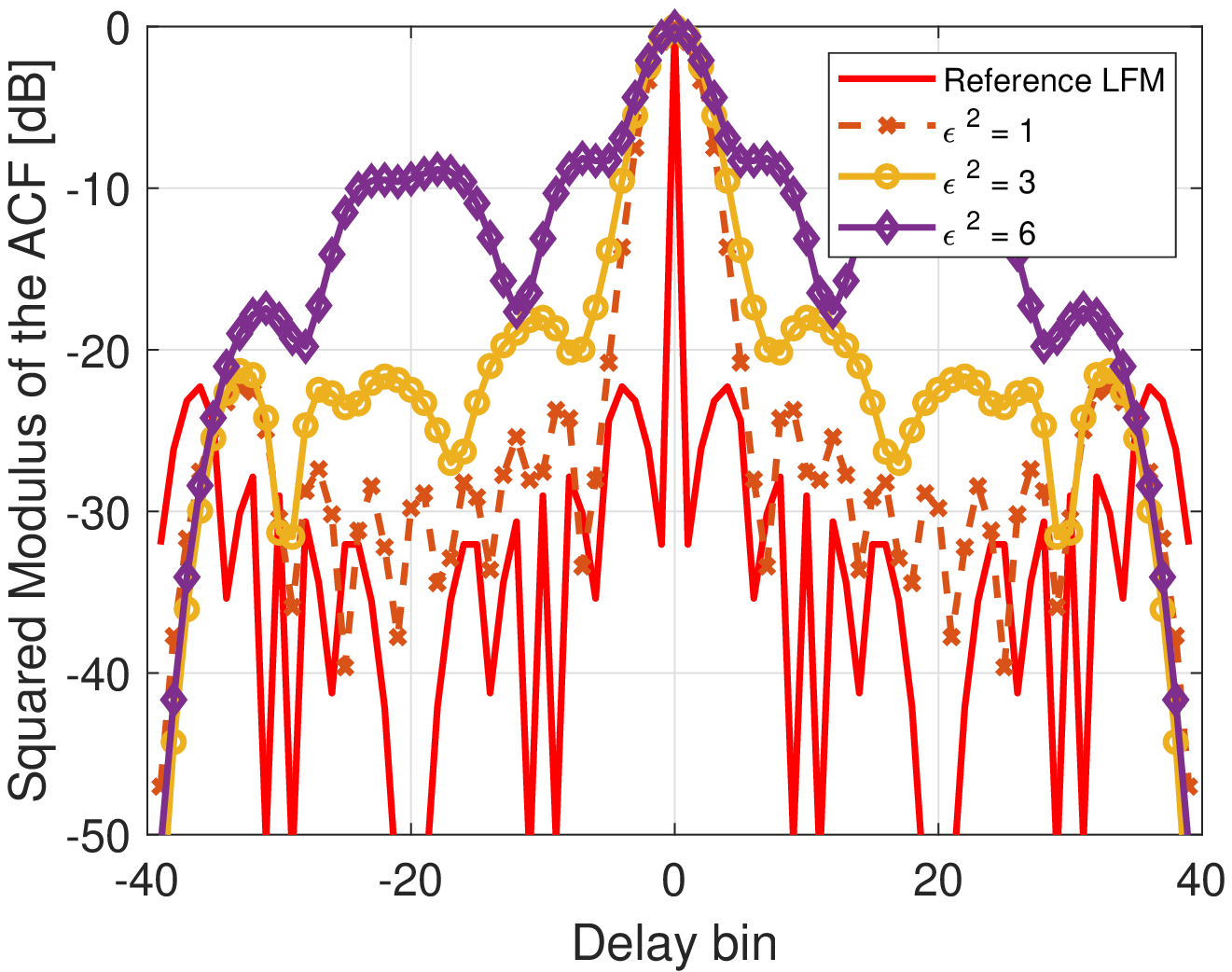}
		\end{minipage}
	}%
	\subfigure[]{
		\begin{minipage}[t]{0.33\linewidth}
			\centering
			\includegraphics[width=0.22\textheight]{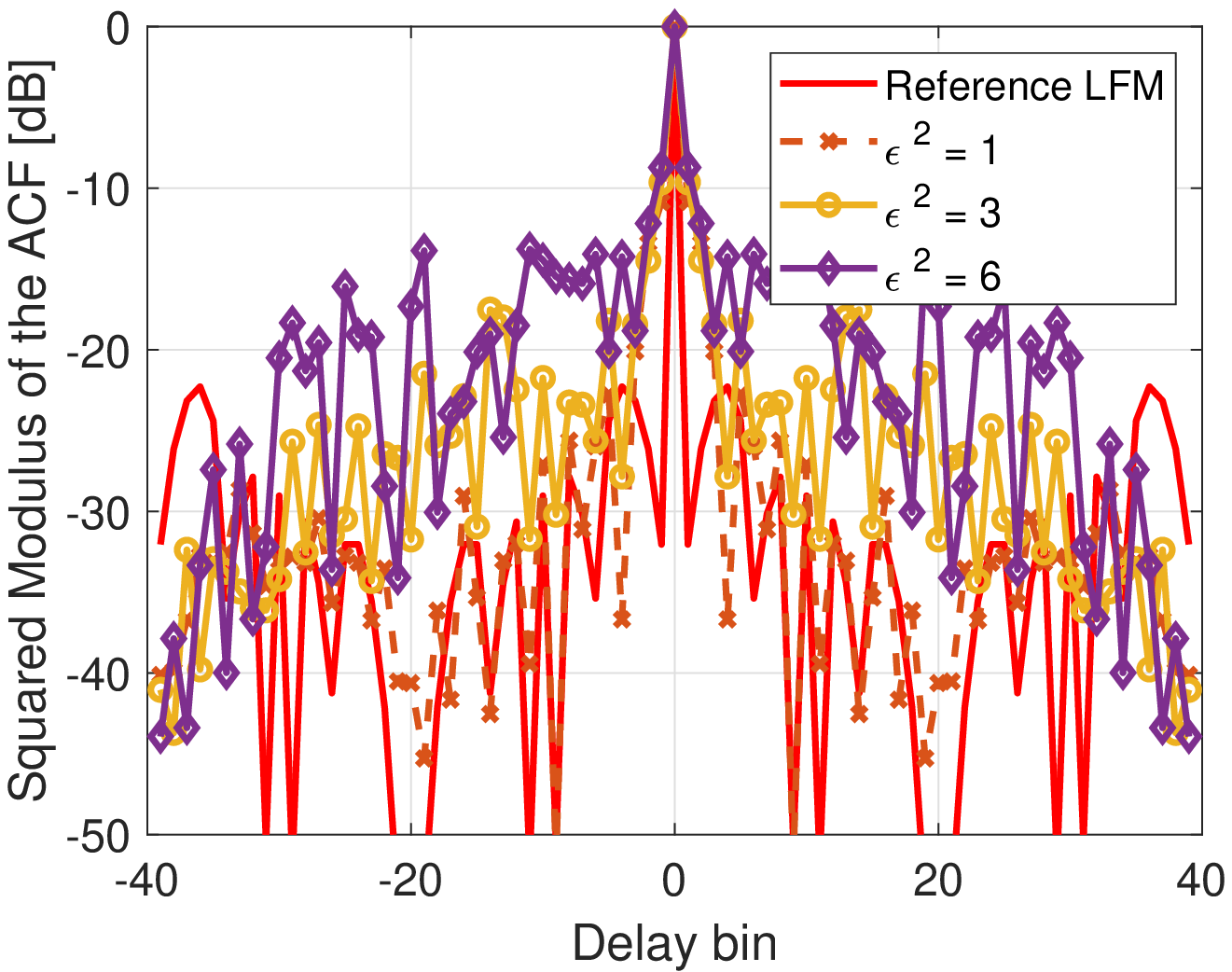}
		\end{minipage}
	}%
	
	\centering
	\caption{The ACF property of the waveforms as compared with the reference LFM. (a) waveform $1$, (b) waveform $2$, (c) waveform $3$, (d) waveform $4$, (e) waveform $5$, and (f) waveform $6$.}
	\label{Fi9}
\end{figure*}

The ACF profiles of the waveforms with different similarity values obtained by the joint design are compared with that of the reference LFM in Fig. \ref{Fi9}.
The results show that the peak sidelobe level (PSL), or range resolution, gets better as the required similarity level decreases, again because of the reduced remaining DoF available to design the waveforms.
This fact suggests a compromise between the SINR and the resolution or PSL characteristics.
It can also be observed that the fluctuations of the designed waveform becomes more pronounced as the similarity value increases. The reason is that for the larger similarity values, the more DoFs are available in the optimization problem.
It should be noted that the waveform of the $5$-th transmitter has a higher PSL and greater range resolution than the other waveforms.
This may be because the second shared band completely occupies its spectrum.

\subsection{Receive Beampattern}
\begin{figure}[t]
	\centering
	\includegraphics[width=0.45\textwidth]{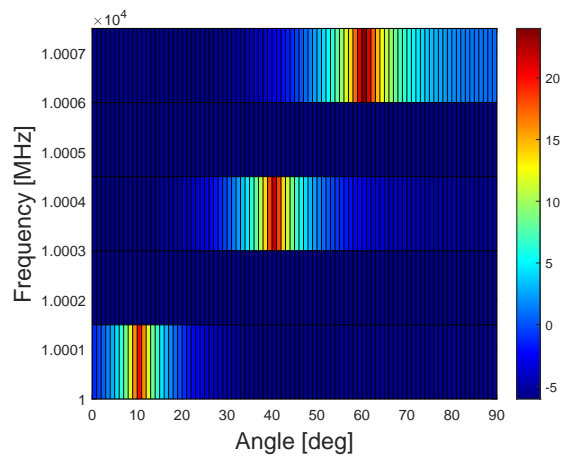}
	\caption{Power spectrum of the interferences distributed in the frequency-angle domain.}
	\label{Fi10}
\end{figure}

\begin{figure}[t]
	\centering
	\includegraphics[width=0.45\textwidth]{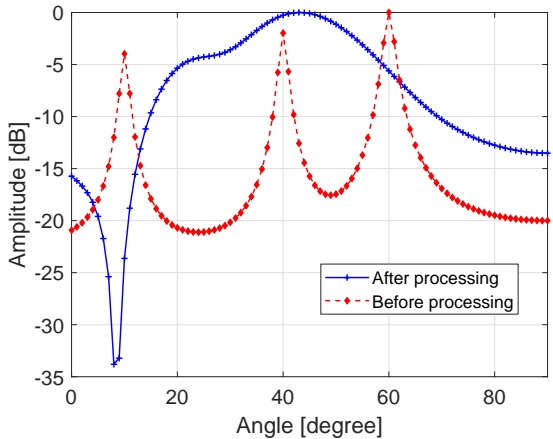}
	\caption{The receive beampattern obtained by jointly design the transmit waveforms and weights.}
	\label{Fi11}
\end{figure}

\begin{figure}[t]
	\centering
	\includegraphics[width=0.45\textwidth]{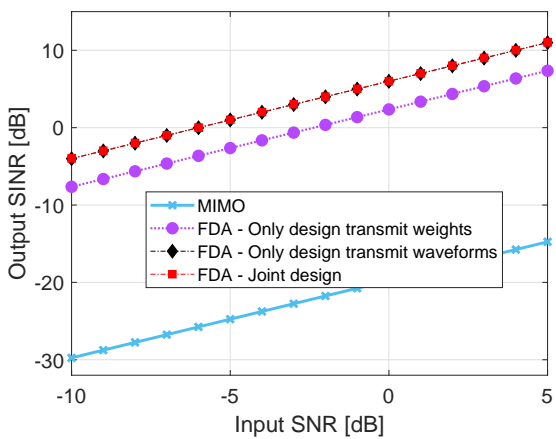}
	\caption{Comparison between the output SINR of MIMO and FDA.}
	\label{Fi12}
\end{figure}

As can be seen in Tab. \ref{symbolss}, interference $2$ enters through the receiver mainlobe pointing to ${{40}^{\text{o}}}$ and others enter through the sidelobe.
Fig. \ref{Fi10} shows the power spectrum of the interferences distributed in the frequency-angle domain, $P\left( {{f}_{m}},\theta  \right)$, calculated by
\begin{equation}
P\left( {{f}_{m}},\theta  \right)={{\left[ {{\mathbf{e}}_{m}}\otimes {{\mathbf{b}}_{\text{R}}}\left( \theta  \right) \right]}^{H}}\mathbf{\bar{Q}}_{i+n}^{-1}\left[ {{\mathbf{e}}_{m}}\otimes {{\mathbf{b}}_{\text{R}}}\left( \theta  \right) \right]
\end{equation}
The resulting receive beampattern ${{P}_{R}}\left( \theta  \right)$ after applying the joint design is displayed in Fig. \ref{Fi11}.
The beampattern is computed as 
\begin{equation}
\begin{aligned}
& {{P}_{R}}\left( \theta  \right) \\ 
& \kern 5pt ={{\left| \mathbf{v}_{opt}^{H}\left\{ {{\mathbf{I}}_{L}}\otimes \left[ {{\mathbf{b}}_{R}}\left( \theta  \right)\otimes \operatorname{diag}\left\{ {{\mathbf{w}}^{*}}\odot {{\mathbf{a}}_{\operatorname{T}}}\left( {{r}_{t}},\theta  \right) \right\} \right] \right\}{{\mathbf{s}}^{*}} \right|}^{2}} \\ 
\end{aligned}
\end{equation}
Although the beampattern is irregular, it can be clearly noticed that nulls are formed at the location of the interference entering through the side lobes, and an energy peak appears at the target location.
Fig. \ref{Fi12} compares the output SINR of FDA and MIMO, where orthogonal LFM is chosen as the transmit waveform for MIMO. 
The high output SINR of FDA demonstrates the effect of interference suppression on the mainlobe.
This advantage is attributed to the additional controllable DoF of FDA in the range dimension.


\section{Conclusions}
\label{sec6}
This paper has addressed the joint design of transmit waveforms and weights for FDA in a spectrally crowded environment.
Employing the multi-channel mixing and low-pass filtering-based FDA receiver, the transmit waveforms are extracted from the signals at the receiver output end. 
Then, considering that the interference exhibits different statistical properties after passing through the FDA receiver, the interference-plus-noise covariance matrix is derived and the multi-channel outputs are synthesized by the MVDR receive weights to suppress the interferences.
In order to maximize the resulting waveform-weight-dependent output SINR and to control the energy distribution over the shared frequency bands, an optimization problem involving a set of energy, bandwidth, and similarity constraints is therefore formulated.
As the resulting problem is non-convex, we develop an iterative SDR algorithm to solve it.
Via simulation, we show that the joint design can not only achieve the expected spectral compatibility, but also obtain the optimal output SINR.
More importantly, the proposed design does not affect the efficient utilization of the controllable DoF of FDA in the range dimension.

\appendix
To prove $(37)$, we divide the constraint into three separate cases on the basis of the frequency band positions occupied by any compatible systems:
\begin{itemize}
	\item[i)] For the case where $f_{l}^{b}$ and $f_{h}^{b}$ are in the same frequency band as the transmit waveform, corresponding to the $f_{l}^{1}$ and $f_{h}^{1}$ in Fig. \ref{Fi2}, i.e., ${{{\tilde{p}}}_{b}}-{{p}_{b}}=0$,  the transmitted energy $E_b$ of FDA in the $b$-th frequency band can be calculated as
	\begin{equation}\label{eq.8}
	{{E}_{b}}=\int_{f_{l}^{b}-{{p}_{b}}\Delta f}^{f_{h}^{b}-{{p}_{b}}\Delta f}{{{\left| {{w}_{{{p}_{b+1}}}}\sum\limits_{l=1}^{L}{\mathbf{S}\left( {{p}_{b}},l \right){{e}^{j2\pi fl}}} \right|}^{2}}\operatorname{d}f},
	\end{equation}
	where $\mathbf{S}=\left[ \mathbf{s}\left( 1 \right),\mathbf{s}\left( 2 \right),...,\mathbf{s}\left( L \right) \right]$ is the transmit waveform matrix, with $\mathbf{S}\left( {{p}_{b}},l \right)$ indicating its $p_b$-th row and $l$-th column element.
	\item[ii)] Similarly, for the case ${{{\tilde{p}}}_{b}}-{{p}_{b}}=1$, corresponding to the $f_{l}^{2}$ and $f_{h}^{2}$ in Fig. \ref{Fi2}, $E_b$ is given as
	\begin{equation}\label{eq.9}
	\begin{aligned}
	& {{E}_{b}}=\int_{f_{l}^{b}-{{p}_{b}}\Delta f}^{\Delta f}{{{\left| {{w}_{{{p}_{b+1}}}}\sum\limits_{l=1}^{L}{\mathbf{S}\left( {{p}_{b}},l \right){{e}^{j2\pi fl}}} \right|}^{2}}\operatorname{d}f} \\ 
	& \kern 17pt +\int_{0}^{f_{h}^{b}-{{{\tilde{p}}}_{b}}\Delta f}{{{\left| {{w}_{{{{\tilde{p}}}_{b}}+1}}\sum\limits_{l=1}^{L}{\mathbf{S}\left( {{{\tilde{p}}}_{b}},l \right){{e}^{j2\pi fl}}} \right|}^{2}}\operatorname{d}f}. \\ 
	\end{aligned}
	\end{equation}
	\item[iii)] Finally, in all other cases, $E_b$ is computed as
	\begin{equation}\label{eq.10}
	\begin{aligned}
	& {{E}_{b}}=\int_{f_{l}^{b}-{{p}_{b}}\Delta f}^{\Delta f}{{{\left| {{w}_{{{p}_{b+1}}}}\sum\limits_{l=1}^{L}{\mathbf{S}\left( {{p}_{b}},l \right){{e}^{j2\pi fl}}} \right|}^{2}}\operatorname{d}f} \\ 
	& \kern 17pt+\int_{0}^{f_{h}^{b}-{{{\tilde{p}}}_{b}}\Delta f}{{{\left| {{w}_{{{{\tilde{p}}}_{b}}+1}}\sum\limits_{l=1}^{L}{\mathbf{S}\left( {{{\tilde{p}}}_{b}},l \right){{e}^{j2\pi fl}}} \right|}^{2}}\operatorname{d}f}\\
	& \kern 17pt+\frac{1}{{{N}_{T}}}\sum\limits_{m={{p}_{b}}+2}^{{{{\tilde{p}}}_{b}}}{{{\left| {{w}_{m}} \right|}^{2}}}. \\ 
	\end{aligned}
	\end{equation} 
\end{itemize}
It is worth noting that \eqref{eq.8}, \eqref{eq.9}, and \eqref{eq.10} can be expressed in a unified and concise form;
taking the third case as an example, \eqref{eq.10} may be rewritten as 
\begin{align}
{E_b} & = w_{{p_b} + 1}^2{\bf{S}}\left( {{p_b}} \right){{\bf K}_{f_l^b}}{{\bf{S}}^H}\left( {{p_b}} \right) \nonumber \\
&\kern 12pt + w_{{{\tilde p}_b} + 1}^2{\bf{S}}\left( {{{\tilde p}_b}} \right){{\bf K}_{f_h^b}}{{\bf{S}}^H}\left( {{{\tilde p}_b}} \right) + \frac{1}{{{N_T}}}\sum\limits_{m = {p_b} + 2}^{{{\tilde p}_b}} {{{\left| {{w_m}} \right|}^2}} \nonumber \\
&= {{\bf{w}}^H}{{{\bf{\tilde I}}}_{{p_b},{{\tilde p}_b}}}\left( {\bf{S}} \right){\bf{w}} \nonumber \\
& = {\bf{s}}_T^H{{\bf{H}}_{{p_b},{{\tilde p}_b}}}{{\bf{s}}_T}
\end{align}
where
\begin{subequations}
	\begin{equation}
	{{\bf{\tilde I}}_{{p_b},{{\tilde p}_b}}}\left( {p,q;{\bf{S}}} \right) = \left\{ {\begin{array}{*{20}{c}}
		{{\bf{S}}\left( {{p_b}} \right){{\bf K}_{f_l^b}}{{\bf{S}}^H}\left( {{p_b}} \right),p = q = {p_b} + 1}\\
		{{\bf{S}}\left( {{{\tilde p}_b}} \right){{\bf K}_{f_h^b}}{{\bf{S}}^H}\left( {{{\tilde p}_b}} \right),p = q = {{\tilde p}_b} + 1}\\
		{\frac{1}{{{N_T}}},p = q = m,m = {p_b} + 2,...,{{\tilde p}_b}}\\
		{0,otherwise}
		\end{array}} \right.,
	\end{equation}
	\begin{equation}
	\kern -12pt {{\bf{H}}_{{p_b},{{\tilde p}_b}}}\left( {p,q;{\bf{w}}} \right) = \left\{ {\begin{array}{*{20}{c}}
		{w_{{p_b} + 1}^2{{\bf K}_{f_l^b}},p = q = {p_b} + 1}\\
		{w_{{{\tilde p}_b} + 1}^2{{\bf K}_{f_h^b}},p = q = {{\tilde p}_b} + 1}\\
		{{{\left| {{w_m}} \right|}^2}{{\bf{I}}_L},p = q \in \left\{ {{p_b} + 2,...,{{\tilde p}_b}} \right\}}\\
		{{{\bf{0}}_L},otherwise}
		\end{array}} \right.
	\end{equation}
\end{subequations}
and
\begin{subequations}
	\begin{equation}
	{{\bf K}_{f_l^b}} = \int_{f_l^b - {p_b}\Delta f}^{\Delta f} {{{{\bf{\tilde e}}}_f}{\bf{\tilde e}}_f^H{\mathop{\rm d}\nolimits} f}
	\end{equation}
	\begin{equation}
	\kern 5pt {{\bf K}_{f_h^b}} = \int_0^{f_h^b - {{\tilde p}_b}\Delta f} {{{{\bf{\tilde e}}}_f}{\bf{\tilde e}}_f^H{\mathop{\rm d}\nolimits} f}
	\end{equation}
	\begin{equation}
	{{\bf K}_{f_l^b,f_h^b}} = \int_{f_l^b - {p_b}\Delta f}^{f_h^b - {p_b}\Delta f} {{{{\bf{\tilde e}}}_f}{\bf{\tilde e}}_f^H{\mathop{\rm d}\nolimits} f}
	\end{equation}
\end{subequations}
Therefore, the FDA spectrally compatibly constraint can be expressed as
\begin{equation}
{E_b} = {{\bf{w}}^H}{{\bf{\tilde I}}_b}\left( {\bf{S}} \right){\bf{w}} = {\bf{s}}_T^H{{\bf{H}}_b}\left( {\bf{w}} \right){{\bf{s}}_T} \le {\eta _b}
\end{equation}
where ${\eta _b}$ represents the acceptable level for the $b$-th compatible systems, and
\begin{subequations}\label{eq:5}
	\begin{equation}\label{eq:51}
	\kern -40pt {{\bf{\tilde I}}_b}\left( {\bf{S}} \right) = \left\{ {\begin{array}{*{20}{c}}
		{{{{\bf{\tilde I}}}_{b,1}}\left( {\bf{S}} \right),{{\tilde p}_b} - {p_b} = 0}\\
		{{{{\bf{\tilde I}}}_{b,2}}\left( {\bf{S}} \right),{{\tilde p}_b} - {p_b} = 1}\\
		{{{{\bf{\tilde I}}}_{b,3}}\left( {\bf{S}} \right),otherwise}
		\end{array}} \right.
	\end{equation}
	\begin{equation}\label{eq:52}
	\kern 4pt {{\bf{\tilde I}}_{b,1}}\left( {{\bf{S}};p,q} \right) = \left\{ {\begin{array}{*{20}{c}}
		{{\bf{S}}\left( {{p_b}} \right){{\bf{{\bf K}}}_{f_l^b,f_h^b}}{{\bf{S}}^H}\left( {{p_b}} \right),p = q = {p_b} + 1}\\
		{0,otherwise}
		\end{array}} \right.
	\end{equation}
	\begin{equation}\label{eq:53}
	\kern -5pt {{\bf{\tilde I}}_{b,2}}\left( {{\bf{S}};p,q} \right) = \left\{ {\begin{array}{*{20}{c}}
		{{\bf{S}}\left( {{p_b}} \right){{\bf K}_{f_l^b}}{{\bf{S}}^H}\left( {{p_b}} \right),p = q = {p_b} + 1}\\
		{{\bf{S}}\left( {{{\tilde p}_b}} \right){{\bf K}_{f_h^b}}{{\bf{S}}^H}\left( {{{\tilde p}_b}} \right),p = q = {{\tilde p}_b} + 1}\\
		{0,otherwise}
		\end{array}} \right.
	\end{equation}
	\begin{equation}\label{eq:54}
	{{\bf{\tilde I}}_{b,3}}\left( {{\bf{S}};p,q} \right) = \left\{ {\begin{array}{*{20}{c}}
		{{\bf{S}}\left( {{p_b}} \right){{\bf K}_{f_l^b}}{{\bf{S}}^H}\left( {{p_b}} \right),p = q = {p_b} + 1}\\
		{{\bf{S}}\left( {{{\tilde p}_b}} \right){{\bf K}_{f_h^b}}{{\bf{S}}^H}\left( {{{\tilde p}_b}} \right),p = q = {{\tilde p}_b} + 1}\\
		{\frac{1}{{{N_T}}},p = q = m,m = {p_b} + 2,...,{{\tilde p}_b}}\\
		{0,otherwise}
		\end{array}} \right.,
	\end{equation}
	\begin{equation}\label{eq:55}
	\kern -37pt {{\bf{H}}_b}\left( {\bf{w}} \right) = \left\{ {\begin{array}{*{20}{c}}
		{{{\bf{H}}_{b,1}}\left( {\bf{w}} \right),{{\tilde p}_b} - {p_b} = 0}\\
		{{{\bf{H}}_{b,2}}\left( {\bf{w}} \right),{{\tilde p}_b} - {p_b} = 1}\\
		{{{\bf{H}}_{b,3}}\left( {\bf{w}} \right),otherwise}
		\end{array}} \right.
	\end{equation}
	\begin{equation}\label{eq:56}
	\kern -24pt {{\bf{H}}_{b,1}}\left( {p,q;{\bf{w}}} \right) = \left\{ {\begin{array}{*{20}{c}}
		{w_{{p_b} + 1}^2{{\bf{{\bf K}}}_{f_l^b}},p = q = {p_b} + 1}\\
		{{{\bf{0}}_L},otherwise}
		\end{array}} \right.
	\end{equation}
	\begin{equation}\label{eq:57}
	\kern -20pt {{\bf{H}}_{b,2}}\left( {p,q;{\bf{w}}} \right) = \left\{ {\begin{array}{*{20}{c}}
		{w_{{p_b} + 1}^2{{\bf{{\bf K}}}_{f_l^b}},p = q = {p_b} + 1}\\
		{w_{{{\tilde p}_b} + 1}^2{{\bf K}_{f_h^b}},p = q = {{\tilde p}_b} + 1}\\
		{{{\bf{0}}_L},otherwise}
		\end{array}} \right.
	\end{equation}
	\begin{equation}\label{eq:58}
	\kern -3.7pt {{\bf{H}}_{b,3}}\left( {p,q;{\bf{w}}} \right) = \left\{ {\begin{array}{*{20}{c}}
		{w_{{p_b} + 1}^2{{\bf K}_{f_l^b}},p = q = {p_b} + 1}\\
		{w_{{{\tilde p}_b} + 1}^2{{\bf K}_{f_h^b}},p = q = {{\tilde p}_b} + 1}\\
		{{{\left| {{w_p}} \right|}^2}{{\bf{I}}_L},p = q \in \left\{ {{p_b} + 2,...,{{\tilde p}_b}} \right\}}\\
		{{{\bf{0}}_L},otherwise}
		\end{array}} \right.
	\end{equation}
\end{subequations}
which completes the proof.

\end{document}